\documentclass[final,onefignum,onetabnum]{siamart190516}
\usepackage{algorithmicx} 
\usepackage{algpseudocode}
\usepackage{esint}
\usepackage{color}

\makeatletter

\numberwithin{equation}{section}
\numberwithin{figure}{section}
\theoremstyle{plain}
\newtheorem{thm}{\protect\theoremname}
  \theoremstyle{remark}

\allowdisplaybreaks

\makeatother

\usepackage{babel}
  \providecommand{\remarkname}{Remark}
\providecommand{\theoremname}{Theorem}

\usepackage{lipsum}
\usepackage{amsfonts}
\usepackage{graphicx}
\usepackage{epstopdf}

\headers{MCMC with teleportation}{M. Lindsey, J. Weare, and A. Zhang}

\title{Ensemble Markov chain Monte Carlo with teleporting walkers\thanks{Submitted to the editors June 4, 2021.
\funding{ M.L. acknowledges support from the National Science Foundation under Award No. 1903031. J.W. acknowledges support from the Advanced Scientific Computing Research Program within the DOE Office of Science through award DE-SC0020427. }}}

\author{Michael Lindsey\thanks{Courant Institute of Mathematical Sciences, New York, New York 10012, United States (\email{lindsey@cims.nyu.edu}, \email{weare@cims.nyu.edu})}
\and Jonathan Weare\footnotemark[2]
\and Anna Zhang\thanks{Stuyvesant High School, New York, New York 10282, United States (\email{azhang03@mit.edu})}}
\usepackage{amsopn}

\ifpdf
\hypersetup{
  pdftitle={MCMC with teleportation},
  pdfauthor={M. Lindsey, J. Weare, and A. Zhang}
}
\fi

\begin{document}
\global\long\def\ve{\varepsilon}
\global\long\def\R{\mathbb{R}}
\global\long\def\Rn{\mathbb{R}^{n}}
\global\long\def\Rd{\mathbb{R}^{d}}
\global\long\def\E{\mathbb{E}}
\global\long\def\P{\mathbb{P}}
\global\long\def\bx{\mathbf{x}}
\global\long\def\vp{\varphi}
\global\long\def\ra{\rightarrow}
\global\long\def\smooth{C^{\infty}}
\global\long\def\Tr{\mathrm{Tr}}
\global\long\def\bra#1{\left\langle #1\right|}
\global\long\def\ket#1{\left|#1\right\rangle }

\newcommand{\ML}[1]{\textcolor{red}{[ML:#1]}}

\maketitle

\begin{abstract}
We introduce an ensemble Markov chain Monte Carlo approach to sampling from a probability density with known likelihood. This method upgrades an underlying Markov chain by allowing an ensemble of such chains to interact via a process in which one chain's state is cloned as another's is deleted. This effective teleportation of states can overcome issues of metastability in the underlying chain, as the scheme enjoys rapid mixing once the modes of the target density have been populated. We derive a mean-field limit for the evolution of the ensemble. We analyze the global and local convergence of this mean-field limit, showing asymptotic convergence independent of the spectral gap of the underlying Markov chain, and moreover we interpret the limiting evolution as a gradient flow. We explain how interaction can be applied selectively to a subset of state variables in order to maintain advantage on very high-dimensional problems. Finally we present the application of our methodology to Bayesian hyperparameter estimation for Gaussian process regression.
\end{abstract}



\section{Introduction}
In practice, the efficiency of a Markov chain Monte Carlo (MCMC) algorithm is often limited by metastability, that is, the need to repeatedly transition between high-probability regions  separated by regions of low probability. Because an MCMC chain is designed to sample each region according to its probability, it will necessarily visit the low-probability region, and therefore also transition between the high-probability regions, only infrequently. In practice, metastability is difficult to address without detailed insights into its origins in the specific problem of interest (e.g., a description of relatively high-probability pathways connecting high-probability regions).  Common approaches to overcoming metastability involve the modification of a general-purpose MCMC algorithm (such as Metropolis--Hastings or Langevin dynamics \cite{liu2001monte,Leimkuhler2015mdbook}) by, e.g., rescaling the log target density by a small factor (as in parallel tempering \cite{liu2001monte,Earl2005partemp}) or stratifying the sampling space (as in umbrella sampling \cite{dinner2020emus,matthews2018us} and related schemes \cite{Chopin2012,webber2020splitting}).

We propose an alternative strategy in which an interaction is introduced between multiple (otherwise independently evolving) chains, specifying the evolution for an ensemble of `walkers.'  At each step of the algorithm, one walker is selected to be duplicated and moved according to some proposal, and another is selected to be removed. If the duplicated and removed walkers are different, we say that a walker has been `teleported.' The scheme involves a Metropolis--Hastings accept-reject step and exactly preserves a specified target density. In the mean-field limit of many walkers, the acceptance probability converges to 1, and our scheme somewhat resembles a resampling strategy \cite{chopin2020smc}.  We identify the mean-field evolution and find that its local convergence to the target is rapid even in cases that would lead to metastabilities in standard single-chain MCMC schemes. In particular, we prove an asymptotic convergence rate for the mean-field evolution that is independent of the spectral gap of the Markov chain used to define the parallel walker evolutions. Moreover, we interpret the mean-field density evolution as a gradient flow \cite{gradFlow} of the $\chi^2$-divergence \cite{divergences} with respect to a metric that resembles the Hellinger distance \cite{divergences}.

A shortcoming of our scheme is that the advantage from interaction tends to decrease as the dimension of the sample space increases relative to the number of walkers.  Fortunately, in this limit our scheme reverts to running independent chains sampling from the target without interaction. Moreover, as we demonstrate, for higher-dimensional sampling problems the interaction we introduce can be restricted to a low-dimensional subspace of state variables.

Ensemble Markov chain Monte Carlo schemes are now implemented in several very popular software packages and have found widespread use on a variety of parameter estimation problems \cite{Foreman2013,zuntz_etal15}.  Most of these schemes use information from the ensemble of chains to address conditioning problems \cite{GilksRoberts:1994:Snooker,Braak:2006:GeneticMCMC,ChristenFox:2010:tWalk,GoodmanWeare:2010:EnsembleMCMC,greengard2015thesis,leimkuhler2018eqn}, i.e., they increase the size of the updates for each chain in directions in which $\pi$ decays relatively slowly, while several articles have emphasized the use of ensemble schemes to avoid  gradient evaluations in traditional optimization and sampling tasks \cite{garbuno2020interactinga,garbuno2020interactingb,pavliotis2021derivativefree,pidstrigach2021affineinvariant}. Recently, studies of the mean-field limit of such schemes have yielded useful new insights \cite{garbuno2020interactinga,garbuno2020interactingb,pidstrigach2021affineinvariant}. Meanwhile, it seems that comparatively few ensemble schemes have been proposed to address slow MCMC convergence due to metastability.
In that our ensemble scheme yields a nonlinear mean-field evolution, it is related  to the `nonlinear' MCMC schemes discussed in \cite{andrieu2007nlmcmc}.  It is more closely related to the ensemble Langevin sampler with birth and death introduced in \cite{lu2019birthdeath}, though that scheme involves additional parameter-dependent approximations.  Similar birth and death dynamics were introduced in \cite{rotskoff2019global} to accelerate training of neural network parameters.

This article is organized as follows. In Section \ref{sec:proposal}, we introduce our ensemble scheme. In Section \ref{sec:continuum}, we formally derive the continuum evolution that emerges in the limit of a large number of walkers, proving global convergence to the target with an asymptotic rate that is independent of the spectral gap of the underlying single-walker Markov chain. We also interpret the evolution as a gradient flow. In Section \ref{sec:subset}, we explain how our scheme can be adapted to introduce interaction only among a subset of state variables. In Section \ref{sec:numerics}, we conclude with numerical experiments. Specifically, we provide a simple illustration of the continuum evolution, and we demonstrate practical performance of our ensemble scheme on Bayesian hyperparameter estimation problems for Gaussian process regression. Under a non-Gaussian measurement noise model, the resulting sampling problem is very high-dimensional and requires us to introduce walker interaction only among a naturally chosen subset of state variables.

\section{Interacting walker proposal} \label{sec:proposal}

Suppose that we are given (up to a possibly unknown normalization) a probability density $\pi(x)$
on a space $X$ and a Markov chain transition density $q(y\,\vert\,x)$
that might serve as a good proposal within a Metropolis--Hastings scheme sampling the target $\pi$.
We want to lift such an approach to an interacting walker approach
on the $N$-fold product space $X^{N}$. Specifically, for a fixed
walker number $N$, we want to sample $\mathbf{x}=(x_{1},\ldots,x_{N})\in X^{N}$
from the probability measure $dM(\mathbf{x})\propto\Pi(\mathbf{x})\,d\mathbf{x}$,
where 
\[
\Pi(\mathbf{x})=\prod_{i=1}^{N}\pi(x_{i}).
\]
 Though the variables $x_{1},\ldots,x_{N}$ are independent with respect
to the joint measure $\Pi$, our chain on $X^{N}$ will not decouple
into $N$ independent chains on $X$.

Note that to any $\mathbf{x}\in X^{N}$ we can associate the empirical
measure $\nu(\mathbf{x})=\frac{1}{N}\sum_{i=1}^{N}\delta_{x_{i}}$.
For bounded continuous $\phi:X\ra\R$ and Borel probability measures
$\nu$, we define $\left\langle \phi,\nu\right\rangle =\E_{\nu}\left[\phi\right]$.
Then we may compute any expectation with respect to the original target
measure $\mu$ as 
\[
\E_{x\sim\mu}\left[\phi(x)\right]=\E_{\mathbf{x}\sim M}\left[\left\langle \phi,\nu(\mathbf{x})\right\rangle \right],
\]
 provided that we can sample from $M$.

Consider the following proposal for an update $\mathbf{x}\ra\mathbf{x}'$.
First uniformly select $j\in\{1,\ldots,N\}$. In our proposal, the
$j$-th sample will be cloned and then moved according to $q$, and
we then sample an index $i$ (possibly equal to $j$) for a sample
to delete from our original set of samples. As such, sample $z\sim q(\,\cdot\,\vert\,x_{j})$.
The index $i$ is then sampled according to the importance weights
\[
w_{i}(\mathbf{x},z):=\frac{q(x_{i}\,\vert\,z)+\sum_{k\neq i}^{N}q(x_{i}\,\vert\,x_{k})}{\pi(x_{i})}\Bigg/Z(\mathbf{x},z),
\]
 where 
\[
Z(\mathbf{x},z):=\sum_{l=1}^{N}\frac{q(x_{l}\,\vert\,z)+\sum_{k\neq l}^{N}q(x_{l}\,\vert\,x_{k})}{\pi(x_{l})}.
\]
 Notice that if $\mathcal{Q}$ is the transition operator on probability
measures induced by $q$, i.e., for a probability measure $\mu$,
\[
\mathcal{Q}\mu(dy)=\int q(dy\,\vert\,x)\,d\mu(x),
\]
then the numerator $q(x_{i}\,\vert\,z)+\sum_{k\neq i}^{N}q(x_{i}\,\vert\,x_{k})$
appearing in the preceding expressions is the density of the measure
$\mathcal{Q}\left[\delta_{z}+\sum_{k\neq i}\delta_{x_{k}}\right]$
evaluated at $x_{i}$. Hence it is improbable to select $i$ for deletion
unless $x_{i}$ is `close' to one of the other samples, i.e., to some $y\in\{x_{1},\ldots,x_{N},z\}\backslash\{x_{i}\}$,
in the sense that $q(x_{i}\,\vert\,y)$ is nonnegligible. Then having
sampled $i$, the proposal is given by $\mathbf{x}'=(x_{k}')$, where
$x_{k}'=x_{k}$ for all $k\neq i$, $x_{i}'=z$. In other words, $x_{i}$
is replaced by $z$ in the proposal.

Supposing that we have generated $\mathbf{x'}$ via the procedure
described above (i.e., so that $i$, $j$, and $z$ are defined as
above), the Metropolis-Hastings acceptance probability can be computed
as 
\[
\min\left(1,\frac{Z(\mathbf{x},z)}{Z(\mathbf{x}',x_{i})}\right).
\]
 (See Appendix \ref{app:acceptance} for a detailed calculation.)
Observe that if none of the walkers are close to one another according
to $q$, i.e., if $q(x_{l}\vert x_{k})\approx0$ for all $k\neq l$
and moreover $q(x_{l}\vert z)\approx0$ for all $l\neq j$, then we
select $i=j$ with high probability, and the acceptance probability
is approximately 
\[
\min\left(1,\frac{q(x_{j}\,\vert\,z)}{\pi(x_{j})}\frac{\pi(z)}{q(z\,\vert\,x_{j})}\right),
\]
 so we default to simply performing a Metropolis update according
to $q$ for the $j$-th sample.

Meanwhile, as we shall discuss in more detail below, one expects $Z(\mathbf{x},z)\approx Z(\mathbf{x}',x_{i})$
when the number of walkers is large. In other words, we expect that the acceptance
probability will approach $1$ as the number of samples is increased,
holding all else constant.

As $N$ increases, one expects a transition from the small-$N$ regime
(in which the walkers are isolated from one another relative to
the proposal kernel) to the large-$N$ regime (in which each walker
has several neighbors relative to the kernel). A curse of dimensionality
enters in that for a fixed proposal kernel that is narrow enough to
yield a nonnegligible acceptance probability, one must take $N$ exponentially
large in the dimension of $X$ in order for each walker to have
several neighbors with respect to this kernel. However, the onset
of the curse is delayed as the proposal is improved; indeed, if $q(y\,\vert\,x)=\pi(y)$,
then by inspection one observes that the importance weights $w_{i}$
are uniform, the acceptance probability is $1$, and the sampler reaches
equilibrium in one step, just as is the case for ordinary MCMC with
a perfect proposal. In practice, we shall observe that the scheme
can still succeed on practical problems in dimensions that are much
too high to treat simply by quadrature.

\section{Large-$N$ limit}\label{sec:continuum}

In this section we consider the scheme introduced in Section \ref{sec:proposal} in the
limit of large $N$. In this limit we will try to identify the empirical measure $\nu=\nu(\mathbf{x})$
with an absolutely continuous measure $d\nu=\rho\ dx$. In this section
we provide a formal derivation of the dynamics that emerge for $\rho$
in this limit. Note that since each update step can only move a single
walker, we only make a change of order $1/N$ to $\nu$. Hence we
want to think of $\Delta t=1/N$.

Notice that if $d\nu\approx\rho\,dx$, we can approximate 
\[
\frac{Z(\mathbf{x},z)}{N^{2}}\approx\frac{Z(\mathbf{x}',x_{i})}{N^{2}}\approx\E_{x\sim\nu}\left[\frac{1}{\pi(x)}\frac{d(\mathcal{Q}\nu)}{dx}\right]=\int\frac{\mathcal{Q}\rho(x)}{\pi(x)}\rho(x)\,dx,
\]
 where we abuse notation slightly to view $\mathcal{Q}$ is an operator
on probability densities as well as measures, i.e., we define $\mathcal{Q}p(x)=\frac{d(\mathcal{Q}\mu)}{dx}$
where $p$ is the density of $\mu$. Note that in particular we
expect the acceptance probability converges to $1$ as $N\ra\infty$.

Consider $\phi:X\ra\R$. Then for $\mathbf{x}$ fixed and $\mathbf{x}'$
(random) obtained by applying one step of our chain to $\mathbf{x}$,
we have 
\[
\E\left(\left\langle \phi,\nu(\mathbf{x}')-\nu(\mathbf{x})\right\rangle \right)\approx\frac{1}{N}\E\left\{ \phi(z)-\phi(x_{i})\right\} 
\]
 for large $N$, since the acceptance probability is approximately
$1$. In the right-hand side, $z$ is sampled by sampling $y\sim\nu(\mathbf{x})$
and then applying one step of $q$ to obtain $z$, i.e., $z$ is sampled
from the density $\mathcal{Q}\rho$, and the index $i$ is sampled
according to the importance weight 
\[
w_{i}(\mathbf{x},z):=\frac{q(x_{i}\,\vert\,z)+\sum_{k\neq i}^{N}q(x_{i}\,\vert\,x_{k})}{\pi(x_{i})}\Bigg/Z(\mathbf{x},z)\approx\frac{\frac{\mathcal{Q}\rho(x_{i})}{\pi(x_{i})}}{\int\frac{\mathcal{Q}\rho(x)}{\pi(x)}\rho(x)\,dx}.
\]
 Hence we can view $y:=x_{i}$ as being approximately sampled from
the importance-weighted density $\frac{1}{Z_{p}}\frac{\mathcal{Q}\rho}{\pi}\rho$,
where $Z_{\rho}:=\int\frac{\mathcal{Q}\rho(x)}{\pi(x)}\rho(x)\,dx$,
and therefore 
\begin{eqnarray*}
\frac{\E\left(\left\langle \phi,\nu(\mathbf{x}')-\nu(\mathbf{x})\right\rangle \right)}{\Delta t} & \approx & \int\phi(z)\,\mathcal{Q}\rho(z)\,dz-\frac{1}{Z_{\rho}}\int\phi(y)\frac{\mathcal{Q}\rho(y)}{\pi(y)}\rho(y)\,dy\\
 & = & \int\phi(x)\,\left[1-\frac{1}{Z_{\rho}}\frac{\rho(x)}{\pi(x)}\right]\mathcal{Q}\rho(x)\,dy.
\end{eqnarray*}
 Now we view $\frac{\E\left(\left\langle \phi,\tilde{\nu}-\nu\right\rangle \right)}{\Delta t}\approx\left\langle \phi,\dot{\rho}\right\rangle _{L^{2}}$,
where we view $\rho=\rho_{t}(x)$ now as time-dependent and take $\dot{\rho}_{t}(x)=\frac{\partial}{\partial t}\rho_{t}(x)$,
so we infer 
\[
\dot{\rho}_{t}(x)=\frac{1}{Z_{\rho_{t}}}\left[Z_{\rho_{t}}-\frac{\rho_{t}(x)}{\pi(x)}\right]\mathcal{Q}\rho_{t}(x).
\]
 For simplicity we shall often write $\rho=\rho_{t}$ and even omit
dependence on $x$, as in 
\begin{equation}
\dot{\rho}=\frac{1}{Z_{\rho}}\left[Z_{\rho}-\frac{\rho}{\pi}\right]\mathcal{Q}\rho.\label{eq:rhodot}
\end{equation}

\subsection{Global convergence analysis}
\label{sec:globalconv}

Our goal in this section is to analyze the convergence of the dynamics
(\ref{eq:rhodot}) to the target density $\pi$. We also highlight
the constrast with the dynamics that arise from considering $N$ independent
Markov chains, each with transition $\tilde{\mathcal{Q}}$ defined
to be the Metropolized version of $\mathcal{Q}$, which satisfies
$\tilde{\mathcal{Q}}\pi=\pi$. These dynamics are specified by 
\begin{equation}
\dot{\rho}=-\left(\mathrm{Id}-\tilde{\mathcal{Q}}\right)\rho,\label{eq:contMetro}
\end{equation}
 as can be verified by an analogous (but simpler) formal calculation.
Equivalently, we have $\dot{\eta}=-\left(\mathrm{Id}-\tilde{\mathcal{Q}}\right)\eta,$
where $\eta:=\rho-\pi$. These dynamics for the error conserve the
constraint $\int\eta\,dx=0$. On the subspace defined by this constraint,
the convergence of the dynamics is linear with rate given by the spectral
gap of $\tilde{\mathcal{Q}}$ \cite{Levin2017book}. Hence the convergence is
slow when the gap is small, which is known to be the case \cite{Lelievre2016acta,Helffer2004metastab},
e.g., for multimodal $\pi$ with local proposals that cannot cross
between modes.

Our ensemble approach cannot `discover' new modes any
faster than would an independent-chain approach. This is intuitive from
the construction, as well as the perspective of Section \ref{sec:gradflow} below, which can
be viewed in part as quantifying the difficulty of expanding the support
of $\rho$. However, once the modes are discovered, the convergence
is potentially much faster, as our local convergence analysis of the continuum limit shall indicate. By contrast, note that for independent walkers, even if all modes are populated by the ensemble, fluctuations in the populations of each mode will dissipate very slowly, leading to very slow convergence.

We approach questions of convergence first by identifying a convenient monotone quantity, defined as a  
Pearson $\chi^{2}$-divergence. Recall that this divergence is defined by the formula \cite{divergences} 
\[
\chi^{2}(\rho_{1}\,\Vert\,\rho_{2}):=\int\left(1-\frac{\rho_{1}(x)}{\rho_{2}(x)}\right)^{2}\rho_{2}(x)\,dx = \int \frac{ \rho_1(x)^2 }{ \rho_2(x)} \,dx - 1.
\]
Then the quantity $\chi^{2}(\pi \,\Vert\,\rho )$ is in fact monotone nonincreasing for the dynamics \eqref{eq:rhodot}, which fact can be verified formally via the computation:
\begin{eqnarray}
\frac{d}{dt}\chi^{2}(\pi\,\Vert\,\rho) & = & \frac{d}{dt}\int\frac{\pi^{2}}{\rho}\,dx \nonumber  \\
 & = & -\int\frac{\pi^{2}}{\rho^{2}}Z_{\rho}^{-1}\left[Z_{\rho}-\frac{\rho}{\pi}\right]\mathcal{Q}\rho\,dx \nonumber \\
 & = & -\left[\int\frac{\pi^{2}}{\rho^{2}}\,\mathcal{Q}\rho\,dx-\frac{\int\frac{\pi}{\rho}\,\mathcal{Q}\rho\,dx}{\int\frac{\rho}{\pi}\,\mathcal{Q}\rho\,dx}\right] \nonumber \\
 & \leq & -\left[\int\frac{\pi^{2}}{\rho^{2}}\,\mathcal{Q}\rho\,dx-\left(\int\frac{\pi}{\rho}\,\mathcal{Q}\rho\,dx\right)^{2}\right] \nonumber \\
 & = & -\int\left[\frac{\pi}{\rho}-\left(\int\frac{\pi}{\rho}\,\mathcal{Q}\rho\,dx\right)\right]^{2}\mathcal{Q}\rho\,dx \nonumber \\
 & = & -\mathrm{Var}_{\mathcal{Q}\rho}(\pi/\rho). \label{eq:monotone}
\end{eqnarray}
 Here the first inequality follows from an application of Jensen's
inequality, and the last expression is interpreted as the variance
of the function $\pi/\rho$ with respect to the density $\mathcal{Q}\rho$. Adopting this notation, observe that $\chi^2 (\pi \, \Vert \, \rho) = \mathrm{Var}_{\rho}(\pi / \rho)$.

Now the quantity $\mathrm{Var}_{\mathcal{Q} \rho}(\pi / \rho)$ is nonnegative and, moreover, equal to zero only if
$\pi=\rho$. Furthermore, $\chi^{2}(\pi\Vert\rho)\geq0$, with equality
if and only if $\pi=\rho$. From monotonicity it should follow that
the dynamics converge to $\pi$. We formalize this claim in the following
theorem, adopting the simplifying assumption that the state space
$X$ is finite. (This assumption simplifies the proof of global-in-time
existence of the dynamics \eqref{eq:rhodot}, but our quantitative arguments rely
on quantities expected to be robust in appropriate limits of infinite
or continuous state spaces.)

\begin{thm} \label{thm:global}
Suppose $X$ is finite, $\mathrm{supp}(\pi)=X$, and $\mathrm{supp}(\mathcal{Q}\rho)=X$
for any probability density $\rho$. Then for any initial probability
density $\rho_{0}$, the dynamics \eqref{eq:rhodot} admit a global-in-time solution
$\rho_{t}$ which converges to $\pi$ as $t\ra\infty$. In fact, 
\begin{equation}
\chi^{2}(\pi\,\Vert\,\rho_{t})\leq e^{-t/\gamma}\chi^{2}(\pi\,\Vert\,\rho_{0}),\label{eq:convbound}
\end{equation}
 where 
\[
\gamma:=\sup_{\rho\ \mathrm{prob.\,dens.}}\left\{ \frac{\mathrm{Var}_{\rho}\left(\pi/\rho\right)}{\mathrm{Var}_{\mathcal{Q} \rho}(\pi/\rho)}\,:\,\chi^{2}(\pi\,\Vert\,\rho)\leq\chi^{2}(\pi\,\Vert\,\rho_{0})\right\} < +\infty.
\]
 In particular, $\gamma=1$ if $\mathcal{Q}=\mathrm{Id}$. In turn we
we have the estimate 
\begin{equation}
\frac{\mathrm{Var}_{\rho}\left(\pi/\rho\right)}{\mathrm{Var}_{\mathcal{Q} \rho}(\pi/\rho)}
\leq
\Vert\rho/\mathcal{Q}\rho\Vert_{\infty}\label{eq:infbound}
\end{equation}
 for all probability densities $\rho$.\end{thm}

The proof is given in Appendix \ref{app:global}.

From (\ref{eq:infbound}) it follows that the asymptotic convergence
rate is at least $\Vert\pi/\mathcal{Q}\pi\Vert_{\infty}^{-1}$. In
particular, if $\mathcal{Q}\pi=\pi$, then the asymptotic convergence
rate is at least $1$ for the $\chi^2$-divergence. We shall see below that in this case, in fact $2$ is the exact asymptotic convergence rate for the $\chi^2$-divergence. We will also see more generally that the lower bound of $\Vert\pi/\mathcal{Q}\pi\Vert_{\infty}^{-1}$ on the asymptotic rate can be improved by a factor of 2. 

Note that $\chi^{2}(\pi\,\Vert\,\rho)=+\infty$ if $\mathrm{supp}(\rho)\neq X$. Therefore the error estimate (\ref{eq:convbound}) is meaningless if the
initial density does not have full support. However, the proof guarantees
that $\mathrm{supp}(\rho_{t})=X$ for any $t>0$. One can in turn
obtain an estimate by viewing some small $t>0$ as the initial time, but
note that the initial $\chi^{2}$-divergence may be extremely large
if, e.g., $\rho_{0}$ puts very little probability on a mode of $\pi$.

Finally, observe that in the case $\mathcal{Q}= \mathrm{Id}$, Theorem \ref{thm:global} furnishes an \emph{a priori} global convergence rate. However, recall that the formal derivation of
the continuum dynamics only makes sense if $\mathcal{Q}$ is nontrivial.
Intuitively, we may think of the case $\mathcal{Q}=\mathrm{Id}$ case
as arising from \emph{first} passing to the large-$N$ limit, \emph{then}
passing to the $\mathcal{Q}\ra\mathrm{Id}$ limit. If $\mathcal{Q}$ is very close
to the identity, we must take $N$ very large to reach the continuum
regime.

\subsection{Asymptotic convergence analysis}
\label{sec:localconv}
It is natural next to linearize the dynamics \eqref{eq:rhodot} about the fixed point $\rho = \pi$ in order to better understand the asymptotic convergence regime. We can rephrase \eqref{eq:rhodot} in terms of the error $\eta=\rho-\pi$ as 
\[
\dot{\eta}=F(\eta)=\frac{1}{Z_{\pi+\eta}}\left[Z_{\pi+\eta}-\frac{\pi+\eta}{\pi}\right]\mathcal{Q}(\pi+\eta),
\]
 where $F$ is suitably defined. In Appendix \ref{app:linearization},
we linearize the dynamics about $\eta=0$ to derive the linearized
system 
\[
\dot{\eta}=\mathcal{J}\eta
\]
 where $\mathcal{J}$ with action defined by 
\[
\mathcal{J}\eta:=DF(0)\eta=  \left(\int \frac{\eta}{\pi}\mathcal{Q}\pi(x)dx - \frac{\eta}{\pi}\right) \mathcal{Q}\pi
\]
 is the suitable Jacobian operator on $S:=\{\eta\,:\,\int\eta\,dx=0\}$.
One can verify by inspection that $\mathcal{J}$ indeed preserves
$S$, at it must because $F$ preserves $S$ as well.

Note that we do not necessarily have $\mathcal{Q}\pi=\pi$
because the transition $\mathcal{Q}$ has not been Metropolized with respect to $\pi$. However, in this natural special case the linearized dynamics simplify tremendously, as the action of Jacobian takes the form
$\mathcal{J}\eta = -\eta$ for any $\eta\in S$.  Because $\chi^2(\pi\Vert \rho)$ has a zero of multiplicity 2 in $\rho$ at the limit point $\rho=\pi,$ this implies that when $\mathcal{Q}\pi=\pi,$ the asymptotic rate of decay of $\chi^2(\pi\Vert \rho)$ is exactly 2.

More generally, the asymptotic convergence rate can be obtained as the smallest eigenvalue of $-\mathcal{J}$ (viewed as an operator
on $S$), provided that the eigenvalues of $\mathcal{J}$ have strictly negative real parts. (In fact we shall see that the eigenvalues are
real and strictly negative.) For simplicity we restrict our attention to the case of finite
state space $X$, so functions can be viewed as finite-dimensional
vectors. In this setting, formal calculations suffice to prove the
following rigorously.
\begin{thm}
\label{thm:continuum}If $X$ is finite and $\mathrm{supp}(\pi)=\mathrm{supp}(\mathcal{Q}\pi)=X$, then the spectrum $\sigma(\mathcal{J})$ of the Jacobian $\mathcal{J}$ satisfies
$\sigma(\mathcal{J})\subset(-\infty,0)$. Let $\alpha = -1/ (\sup{\sigma(\mathcal{J})})$. Then  $\alpha \leq \Vert\pi/\mathcal{Q}\pi\Vert_{\infty}.$
Given a choice of norm and an initial condition $\rho_0$ sufficiently close to $\pi$, for any $\varepsilon > 0$ there
exists $C>0$ such that the dynamics (\ref{eq:rhodot}) converge to
$\pi$ with $\Vert\rho_t -\pi\Vert \leq Ce^{-t/(\alpha + \varepsilon)}$.\end{thm}

The proof is given in Appendix~\ref{app:linearization}.

Because of the multiplicity of the zero $\rho=\pi$ of $\chi^2(\pi\Vert \rho)$, Theorem~\ref{thm:continuum} implies a lower bound of $2 \Vert \pi / \mathcal{Q} \pi \Vert_\infty^{-1} $ on the asymptotic rate of decay for $\chi^2 (\pi \, \Vert \, \rho)$, twice the asymptotic rate of decay guaranteed by Theorem~\ref{thm:global}.

\subsection{Gradient flow structure}
\label{sec:gradflow}

The dynamics (\ref{eq:rhodot}) admit characterization as a gradient
flow \cite{gradFlow}, as we shall now demonstrate formally.

As a warmup we consider a special case: \emph{after} taking this large-$N$
limit, consider then taking the limit $\mathcal{Q}\ra\mathrm{Id}$,
i.e., the limit in which the proposal is trivial. We obtain the equation
\[
\dot{\rho}=\frac{1}{Z_{\rho}}\left[Z_{\rho}-\frac{\rho}{\pi}\right]\rho.
\]
 Observe that the fixed points of the dynamics are those $\rho$ such
that $\rho\,\vert_{\mathrm{supp}(\rho)}\propto\pi\,\vert_{\mathrm{supp}(\rho)}$,
and moreover, the dynamics cannot expand the support of $\rho$. In
fact, if $\mathrm{supp}(\rho_{t})=\mathrm{supp}(\pi)$ at any time
$t$, we will see that $\rho_{t}\ra\pi$ in a suitable sense as $t\ra\infty$.

To make matters simpler, consider a monotonic time-change $\tau=\tau(t)$,
with inverse $t=t(\tau)$, such that $\frac{\partial t}{\partial\tau}=Z_{\rho_{t}}$.
Then identifying $\rho=\rho_{t(\tau)}$ (by a further slight abuse
of notation), we have 
\begin{equation}
\partial_{\tau}\rho=\left[Z_{\rho}-\frac{\rho}{\pi}\right]\rho=\left[1-\frac{\rho}{\pi}\right]\rho+C_{\rho}\,\rho,\label{eq:timechange}
\end{equation}
 where $C_{\rho}:=Z_{\rho}-1$. Notice that $C_{\rho}$ is the unique
choice of constant to ensure that the dynamics conserve total probability.

 We claim that (\ref{eq:timechange}) is a the gradient flow of the
energy $E(\rho):=\frac{1}{8}\chi^{2}(\rho\,\Vert\,\pi)$ with respect
to the metric on the space of probability measures induced by the
Hellinger distance $H$ \cite{divergences}, whose square is defined by: 
\[
H^{2}(\rho_{1},\rho_{2})=\frac{1}{2}\int\left(\sqrt{\rho_{1}(x)}-\sqrt{\rho_{2}(x)}\right)^{2}\,dx.
\]
 Notice that the pointwise square root maps probability densities
to the unit sphere (i.e., $L^{2}$-normalized densities), and the
Hellinger distance is the Euclidean distance pulled back via this
map. Notice further that expanding the support constitutes an infinitely
steep move according to the Hellinger distance (owing to the fact
that $\frac{d}{dq}\vert_{q=0}\sqrt{q}=+\infty$), consistent with
the fact that the dynamics for trivial $q$ cannot expand the support.) Finally, observe that in the energy $E(\rho)$, the target density $\pi$ now appears in the \emph{second}---not the first---slot of the $\chi^2$-divergence, by contrast to the expressions considered in our earlier convergence arguments.

Now the metric only matters (for the purpose of defining a gradient
flow) up to its local expansion up to second order 
\[
H^{2}(\rho+\Delta\rho,\rho)=\frac{1}{4}\int\frac{\Delta\rho(x)^{2}}{\rho(x)}\,dx+...
\]
 Hence $H$ defines a diagonal Riemannian metric on the space of probability
measures. In the finite-dimensional setting, i.e., if $\rho=(\rho_{i})$
is a density on a finite state space, then the metric is given by
$\delta_{ij}/\rho_{i}\ d\rho^{i}\,d\rho^{j}$. Generally we will write
our Riemannian metric as $\frac{\delta(x,y)}{\rho(x)}\,d\rho(x)\,d\rho(y)$.

Then the corresponding gradient flow is defined \cite{gradFlow} by $\partial_{\tau}\rho=\lim_{\ve\ra0^{+}}\frac{\rho_{\ve}-\rho}{\ve}$,
where we in turn define 
\begin{equation}
\rho_{\ve}:=\underset{\tilde{\rho}\in\mathcal{P}(X)}{\mbox{argmin}}\left\{ E(\tilde{\rho})+\frac{1}{2\ve}H^{2}(\tilde{\rho},\rho)\right\} ,\label{eq:argmin}
\end{equation}
and where we allow $\mathcal{P}(X)$ to denote the space of probability
densities on $X$. We formally verify in Appendix \ref{app:gradflow}
that this prescription recovers the dynamics (\ref{eq:timechange}).

By simple modifications to our calculations, we observe that instead
of introducing the time-change, we could have considered the original
dynamics as a gradient flow of $\chi^{2}(\rho\,\Vert\,\pi)$ with respect
to the Riemannian metric 
\[
\frac{8Z_{\rho}\delta(x,y)}{\rho(x)}\,d\rho(x)\,d\rho(y).
\]
 However, to our knowledge this metric does not coincide with any
named metric.

Finally, it follows from simple substitutions in our computations
that the evolution (\ref{eq:rhodot}) for general $\mathcal{Q}$ can
be retrieved as the gradient flow of $\chi^{2}(\rho\,\Vert\,\pi)$ with
respect to the Riemannian metric 
\[
\frac{8Z_{\rho}\delta(x,y)}{\mathcal{Q}\rho(x)}\,d\rho(x)\,d\rho(y),
\]
 which itself depends on the transition operator $\mathcal{Q}$. Hence
in particular the $\chi^{2}$-divergence is monotonically decreasing
on the trajectory. Meanwhile, one notes via inspection of (\ref{eq:rhodot})
that the only fixed points of the dynamics are those $\rho$ such
that $\rho\,\vert_{\mathrm{supp}(\mathcal{Q}\rho)}\propto\pi\,\vert_{\mathrm{supp}(\mathcal{Q}\rho)}$.
If one assumes that $\mathrm{supp}(\mathcal{Q}\rho)=X$ for any $\rho$,
then it follows that the only fixed point is $\rho=\pi$.

\section{Interaction for a subset of variables}\label{sec:subset}

For very high-dimensional problems, the aforementioned curse of dimensionality
reduces the scheme outlined in Section \ref{sec:proposal} to effectively running $N$
independent Markov chains. However, we can modify our scheme to treat
\emph{some} of the state dimensions by an interacting walker scheme and
the rest by ordinary independent Markov chains. In practice, such
a modification may be applicable if there is, e.g., multimodality
with respect to some subset of the variables and fast mixing with
respect to the others. In fact, one might only be interested in expectations
with respect to the former subset, in which case the others may be
viewed as `nuisance variables.'

Concretely, suppose that we can split $X=X^{(1)}\times X^{(2)}$ and
write $x=(u,v)\in X$ where $u\in X^{(1)},v\in X^{(2)}$. We will
sample elements 
\[\mathbf{x}=(\mathbf{u},\mathbf{v})=(u_{1},\ldots,u_{N},v_{1},\ldots,v_{N})\in\left(X^{(1)}\right)^{N}\times\left(X^{(2)}\right)^{N}\]
according to the density 
\[
\Pi(\mathbf{u},\mathbf{v})=\prod_{i=1}^{N}\pi(u_{i},v_{i}).
\]

We will do so be alternating between two sampling stages. First, viewing
$\mathbf{v}$ as fixed, we will construct a Markov chain on $\mathbf{u}$
that conserves the distribution $\Pi(\,\cdot\,,\mathbf{v})\propto\prod_{i=1}^{n}\pi_{v_{i}}^{(1)}(\,\cdot\,)$,
where $\pi_{v_{i}}^{(1)}(u_{i}):=\pi(u_{i},v_{i})$. This chain will
correlate the samples $u_{1},\ldots,u_{N}$, and we will run it for
one step. Then for the second stage, we independently propose updates
$v_{i}'$ for the $v_{i}$ according to some kernel $r(\,\cdot\,\vert\,v_{i})$
on $X^{(2)}$ and accept or reject according to the Metropolis-Hastings
rule for the density proportional to $\pi_{u_{i}}^{(2)}(\,\cdot\,):=\pi(u_{i},\,\cdot\,)$.
This step can be trivially parallelized over the $i$ and can in fact
be repeated many times before returning to the first stage.

Now we turn to a more detailed description of the interacting
stage, which proceeds by analogy to the scheme considered above, subject to
a few necessary modifications. Again we sample $j\in\{1,\ldots,N\}$
uniformly, then sample $z\sim q(\,\cdot\,\vert\,u_{j})$, where $q$
is some transition kernel on $X^{(1)}$. Next we sample $i$ according
to the importance weights 
\[
w_{\mathbf{v},i}(\mathbf{u},z):=\pi_{v_{i}}^{(1)}(z)\frac{q(u_{i}\,\vert\,z)+\sum_{k\neq i}^{N}q(u_{i}\,\vert\,u_{k})}{\pi_{v_{i}}^{(1)}(u_{i})}\Bigg/Z_{\mathbf{v}}(\mathbf{u},z),
\]
 where 
\[
Z_{\mathbf{v}}(\mathbf{u},z):=\sum_{l=1}^{N}\pi_{v_{l}}^{(1)}(z)\frac{q(u_{l}\,\vert\,z)+\sum_{k\neq l}^{N}q(u_{l}\,\vert\,u_{k})}{\pi_{v_{l}}^{(1)}(u_{l})}.
\]
 Relative to our previous importance weights, we have included a factor
of $\pi_{v_{i}}^{(1)}(z)$. In the special case where $X=X^{(1)}$
(i.e., the case considered earlier), such a factor does not affect
the importance weights since it simply acts as a scalar multiplier
independent of $i$. However, in the more general case, the factor
ensures that the scheme is independent of the \emph{relative} normalizations
of the $\pi_{v_{i}}^{(1)}$. As above, having sampled $i$, the proposal
is given by $\mathbf{u}'=(u_{k}')$, where $u_{k}'=u_{k}$ for all
$k\neq i$, $u_{i}'=z$. By analogous computations we find that the
acceptance probability is 
\[
\min\left(1,\frac{\pi_{v_{i}}^{(1)}(u_{i})}{\pi_{v_{i}}^{(1)}(z)}\frac{Z_{\mathbf{v}}(\mathbf{u},z)}{Z_{\mathbf{v}}(\mathbf{u}',u_{i})}\right).
\]

\section{Numerical experiments}\label{sec:numerics}
In this section we provide numerical illustrations of our ensemble scheme and its continuum dynamics \eqref{eq:rhodot} in the large-$N$ limit. First, in Section \ref{sec:continuum_num}, we simulate \eqref{eq:rhodot} and contrast with the dynamics \eqref{eq:contMetro} that arise from the large-$N$ limit for independent (non-interacting) Markov chains.

Then in Section \ref{sec:gpr} we demonstrate the application of the ensemble scheme itself to Bayesian hyperparameter estimation problems in Gaussian process regression. Under a Gaussian measurement noise model, the resulting sampling problems are low-dimensional enough to approach with the fully interacting scheme of Section \ref{sec:proposal}. With non-Gaussian measurement noise, we are led to a very high-dimensional sampling problem for which it is natural to consider the scheme of Section \ref{sec:subset} which introduces interaction for a subset of variables.

\subsection{Continuum dynamics}\label{sec:continuum_num}

We illustrate the continuum dynamics (\ref{eq:rhodot}) 
with a simple numerical simulation. Consider the case $X=\R$ with
the double-well probability density 
\[
\pi(x)=e^{-\beta(x^{4}-x^{2})},
\]
 where $\beta>0$ is an inverse temperature parameter. Note that $\pi$
has modes at $x=\pm\sqrt{1/2}$. We consider the Gaussian proposal
\[
q(x\,\vert\,z)\propto e^{-(x-z)^{2}/2\sigma^{2}},
\]
 where $\sigma>0$ is a parameter controlling the standard deviation
of the proposal. We will compare the dynamics (\ref{eq:rhodot}) against
the continuum dynamics (\ref{eq:contMetro}) for the Metropolized
chain. We refer to these two alternatives respectively as the nonlinear and linear dynamics.

As our initial condition $\rho_{0}$ we consider a mixture of two
Gaussians centered at the modes of $\pi$, 
\[
\rho_{0}(x)\propto\frac{9}{10}e^{-10\cdot\beta\left(x+\sqrt{1/2}\right)^{2}}+\frac{1}{10}e^{-10\cdot\beta\left(x-\sqrt{1/2}\right)^{2}},
\]
 placing 90\% probability on the left mode and 10\% on the right,
with standard deviations tuned to remain within the effective support
of $\pi$.

As a proxy for measuring the convergence of $\rho$ to $\pi$ as $t\ra\infty$,
we simply estimate 
\[
E(t)=\frac{1}{2}-\int_{0}^{\infty}\rho_{t}(x)\,dx,
\]
 where the integral measures the probability according to $\rho$
of a nonnegative sample, which approaches $\frac{1}{2}$ from below
according to either choice of dynamics, as probability is balanced
between the two modes.

\begin{figure}
\centering{}\label{fig:conv}\includegraphics[bb=60bp 0bp 500bp 420bp,clip,scale=0.41]{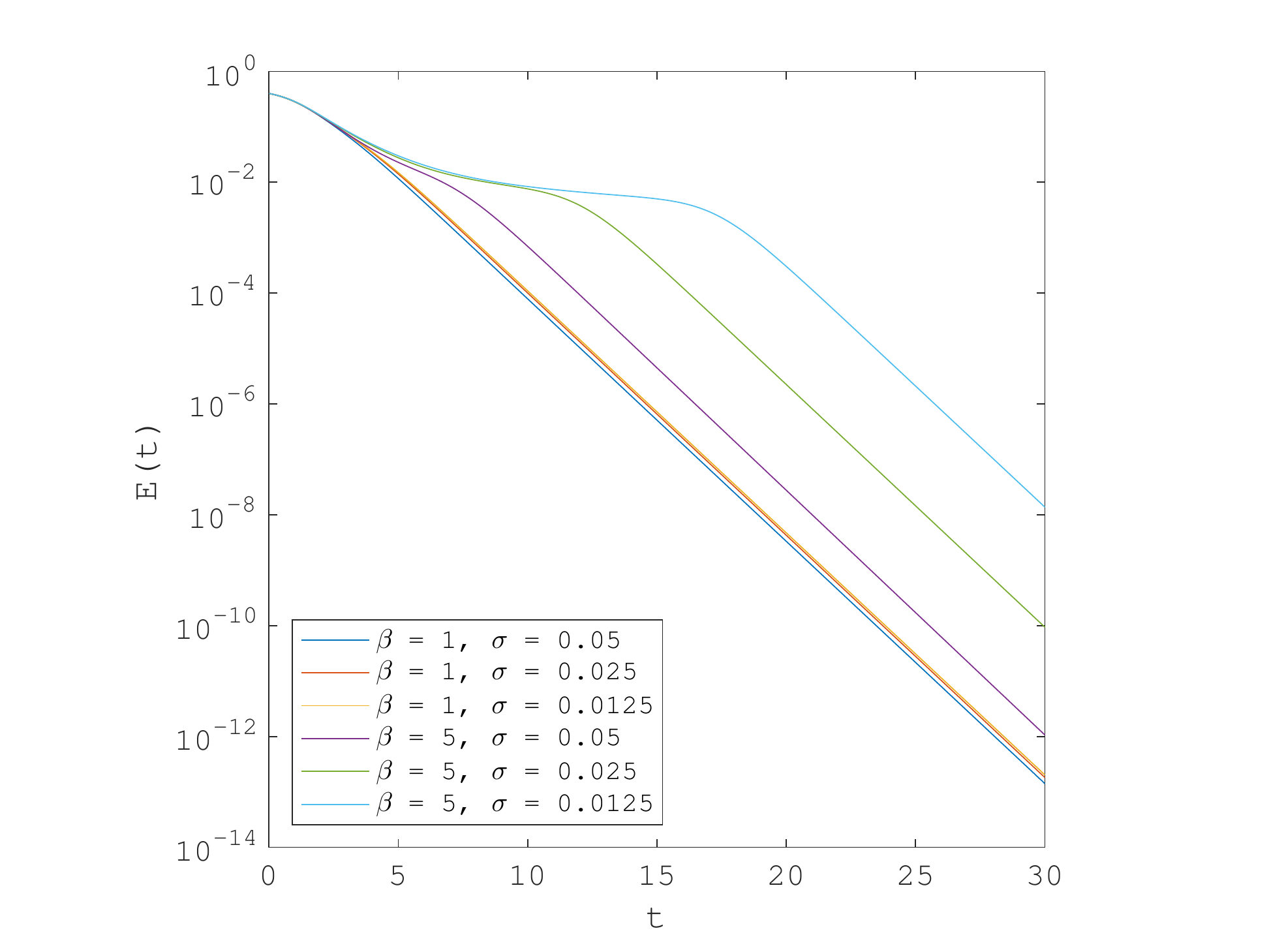}\includegraphics[bb=60bp 0bp 500bp 420bp,clip,scale=0.41]{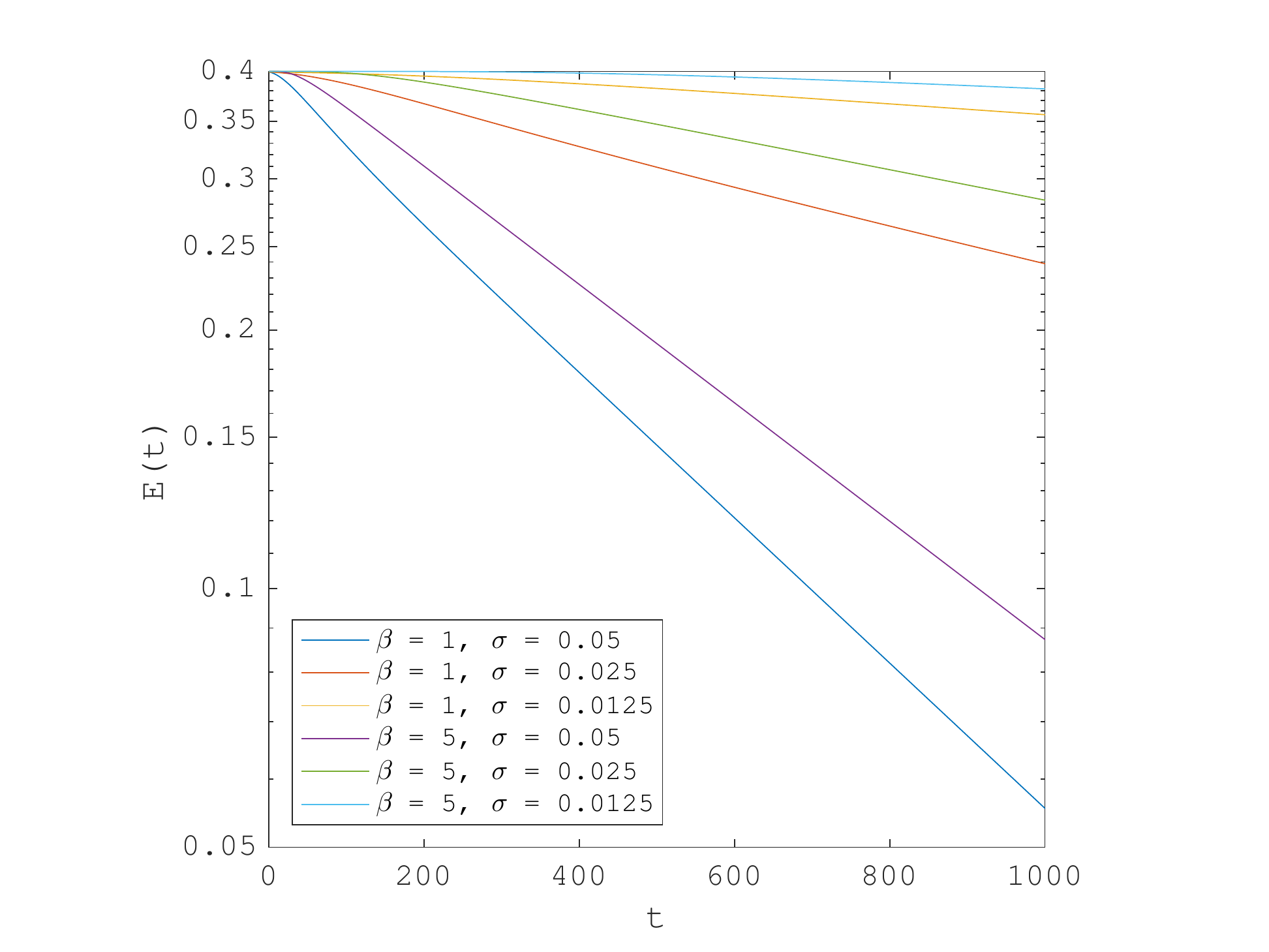}\caption{$E(t)$ for the nonlinear dynamics (\ref{eq:rhodot}) (left) and continuum Metropolis
dynamics (\ref{eq:contMetro}) (right), for several different values
of $\beta,\sigma$. Note the different horizontal and vertical axis
scales at left and right.}
\end{figure}

We discretize both (\ref{eq:rhodot}) and (\ref{eq:contMetro}) with
a simple forward Euler scheme with time-step $\Delta t=0.01$ on an
evenly spaced discretization of the interval $[-2,2]$ with $1000$
points, sufficient for an accurate representation of the dynamics.
We illustrate the convergence $E(t)\ra0$ of both dynamics in Figure
\ref{fig:conv}.

 Observe that within both schemes we observe linear convergence of
the form 
\[
E(t)=Ce^{-t/\alpha}.
\]
 Note that $\alpha$ does not depend noticeably on $\beta,\sigma$
for the nonlinear dynamics (\ref{eq:rhodot}) (and in fact is numerically
close to $1$, consistent with Theorem \ref{thm:continuum}). Meanwhile,
as expected, $\alpha$ depends dramatically on $\beta,\sigma$ for
the continuum Metropolis dynamics (\ref{eq:contMetro}).

\begin{figure}
\begin{centering}
\label{fig:movie}\includegraphics[bb=100bp 35bp 480bp 420bp,clip,scale=0.19]{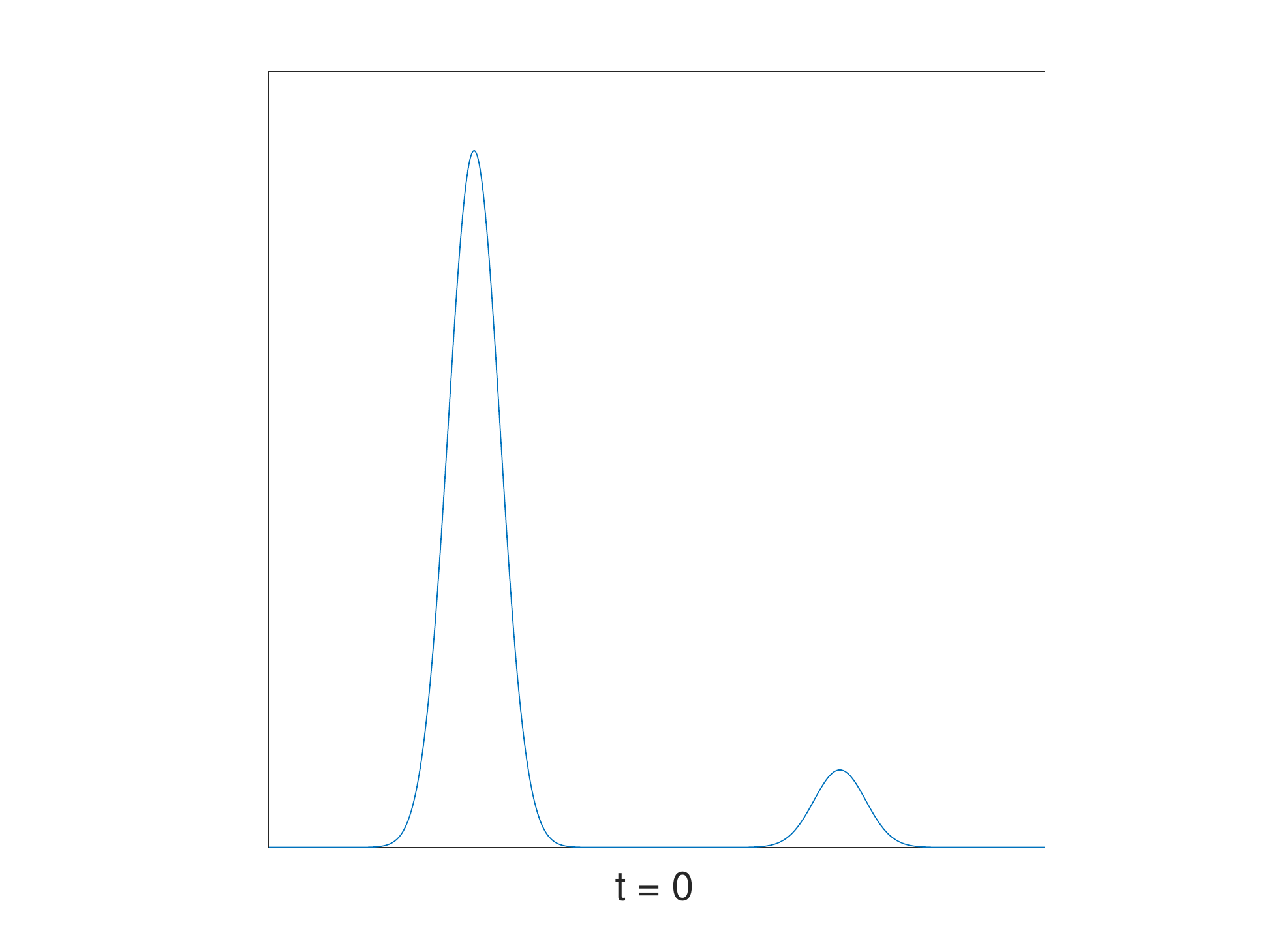}\includegraphics[bb=100bp 35bp 480bp 420bp,clip,scale=0.19]{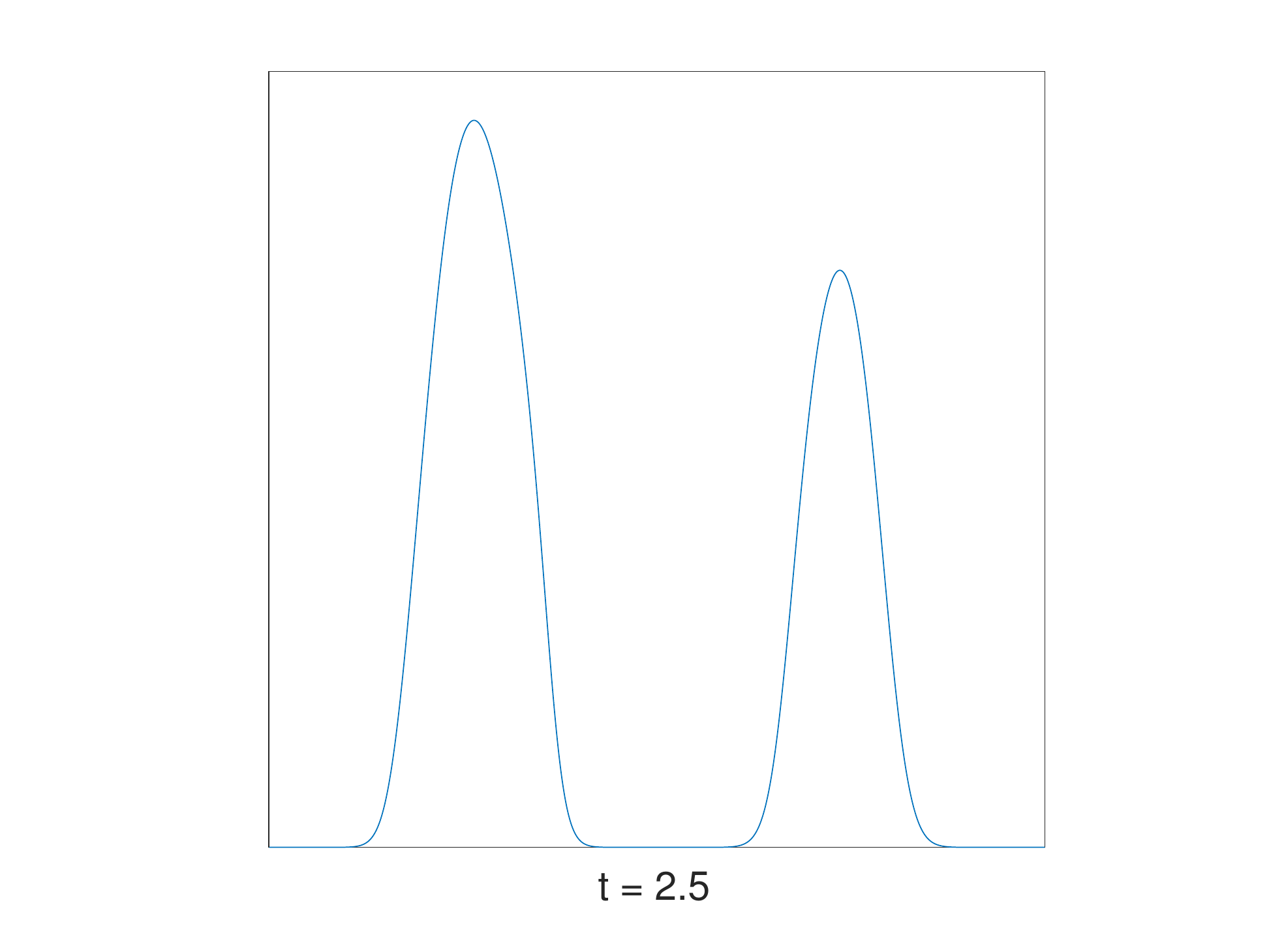}\includegraphics[bb=100bp 35bp 480bp 420bp,clip,scale=0.19]{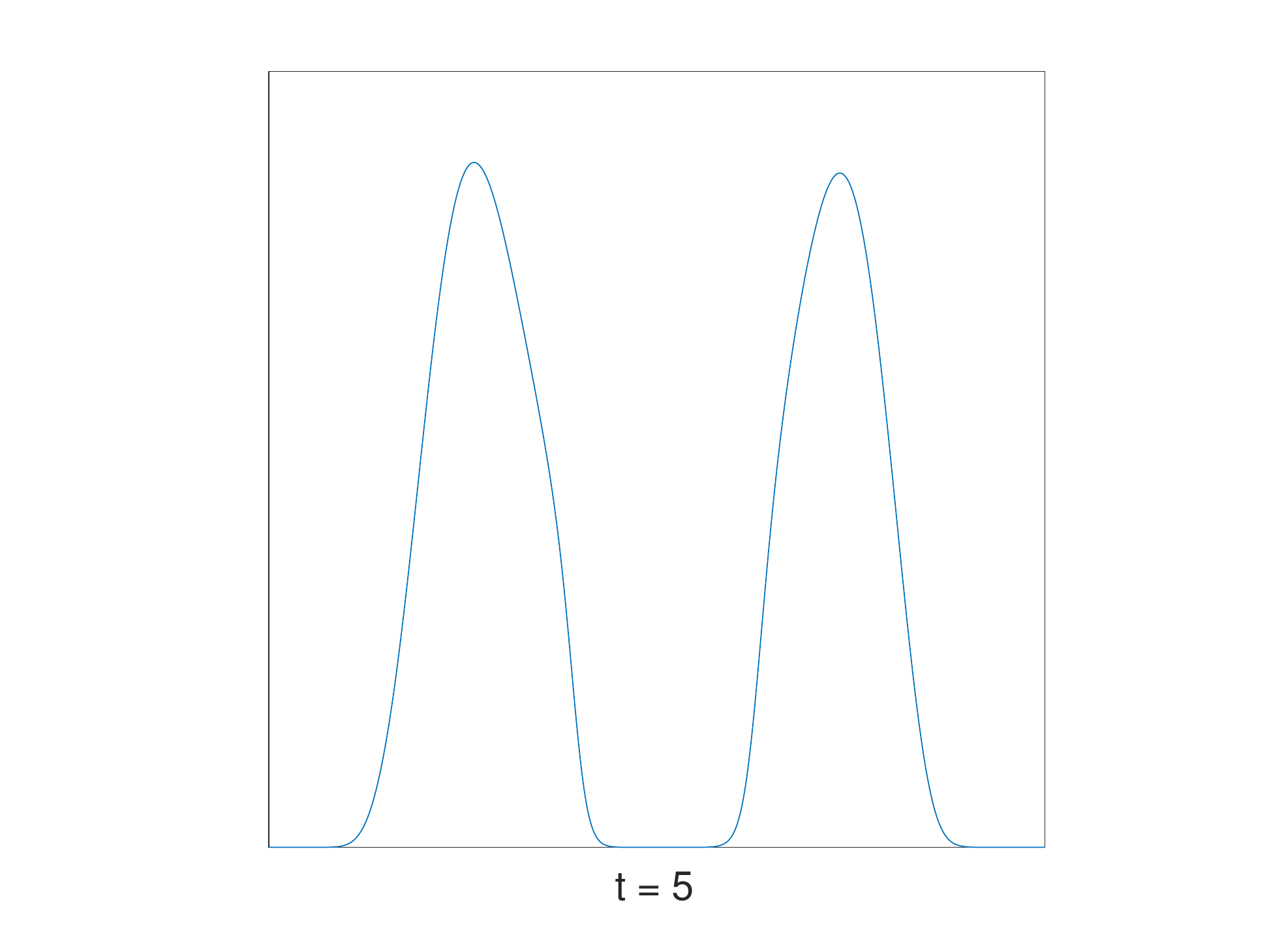}\includegraphics[bb=100bp 35bp 480bp 420bp,clip,scale=0.19]{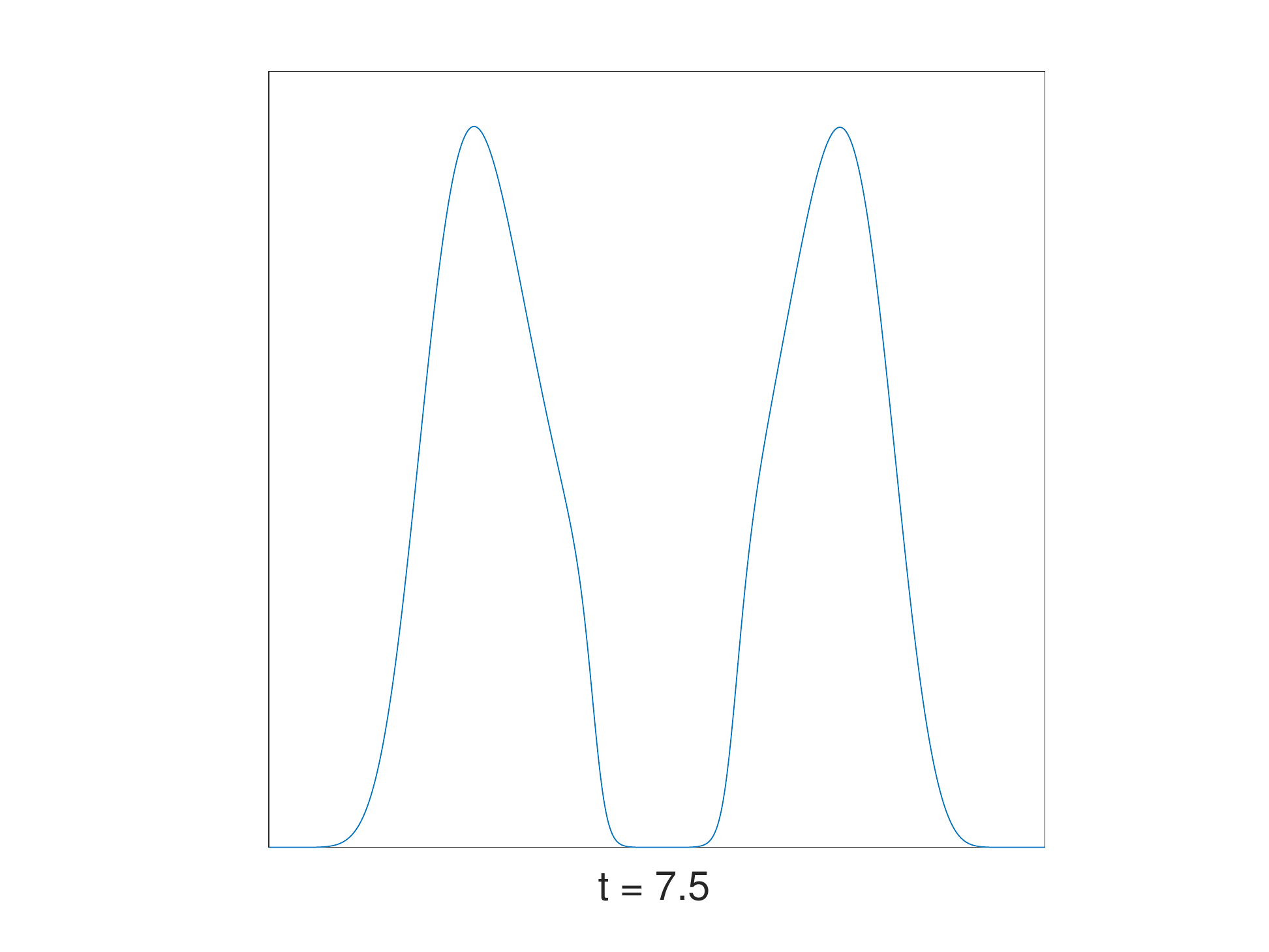}\includegraphics[bb=100bp 35bp 480bp 420bp,clip,scale=0.19]{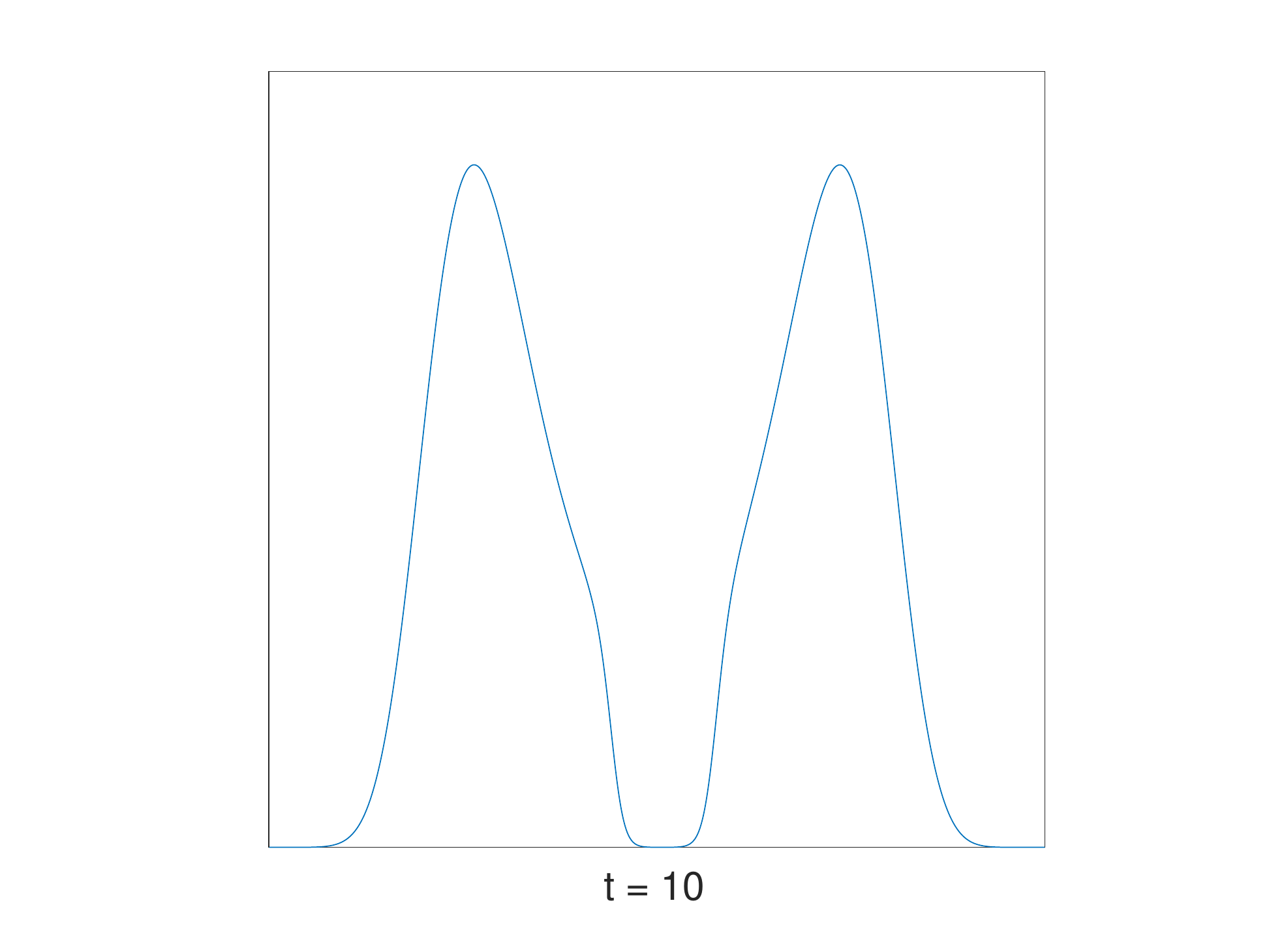}
\par\end{centering}

\centering{}\includegraphics[bb=100bp 35bp 480bp 420bp,clip,scale=0.19]{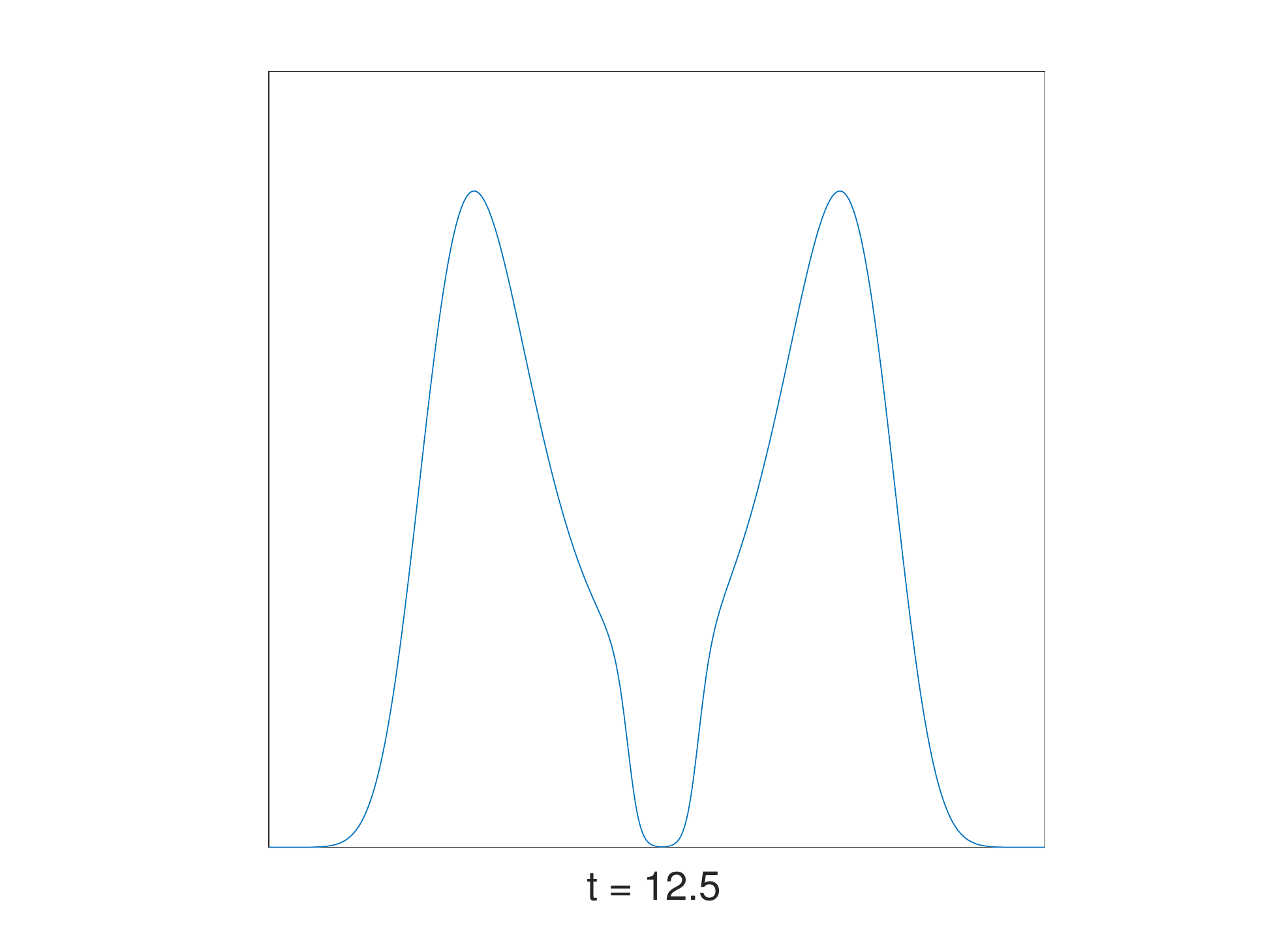}\includegraphics[bb=100bp 35bp 480bp 420bp,clip,scale=0.19]{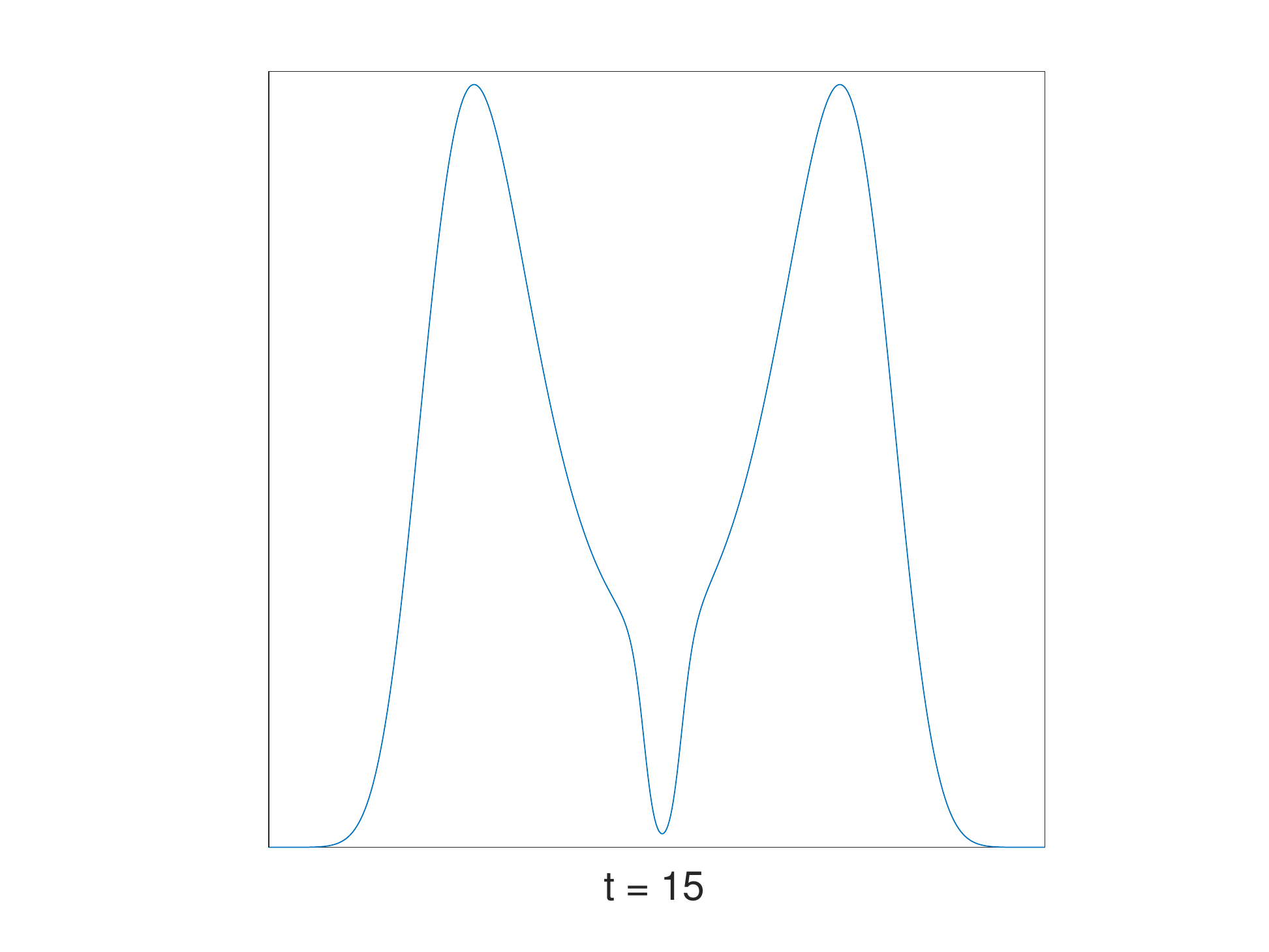}\includegraphics[bb=100bp 35bp 480bp 420bp,clip,scale=0.19]{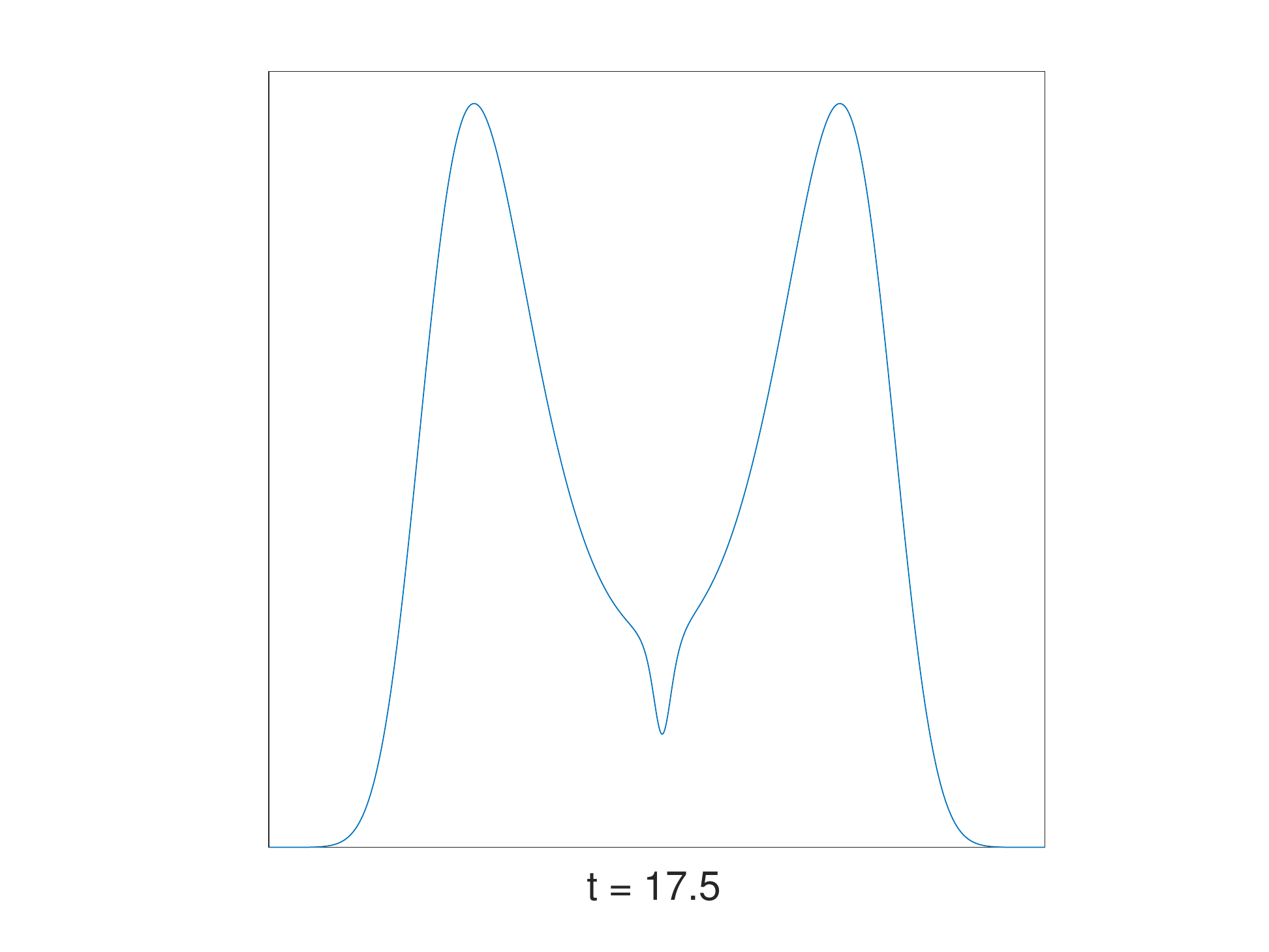}\includegraphics[bb=100bp 35bp 480bp 420bp,clip,scale=0.19]{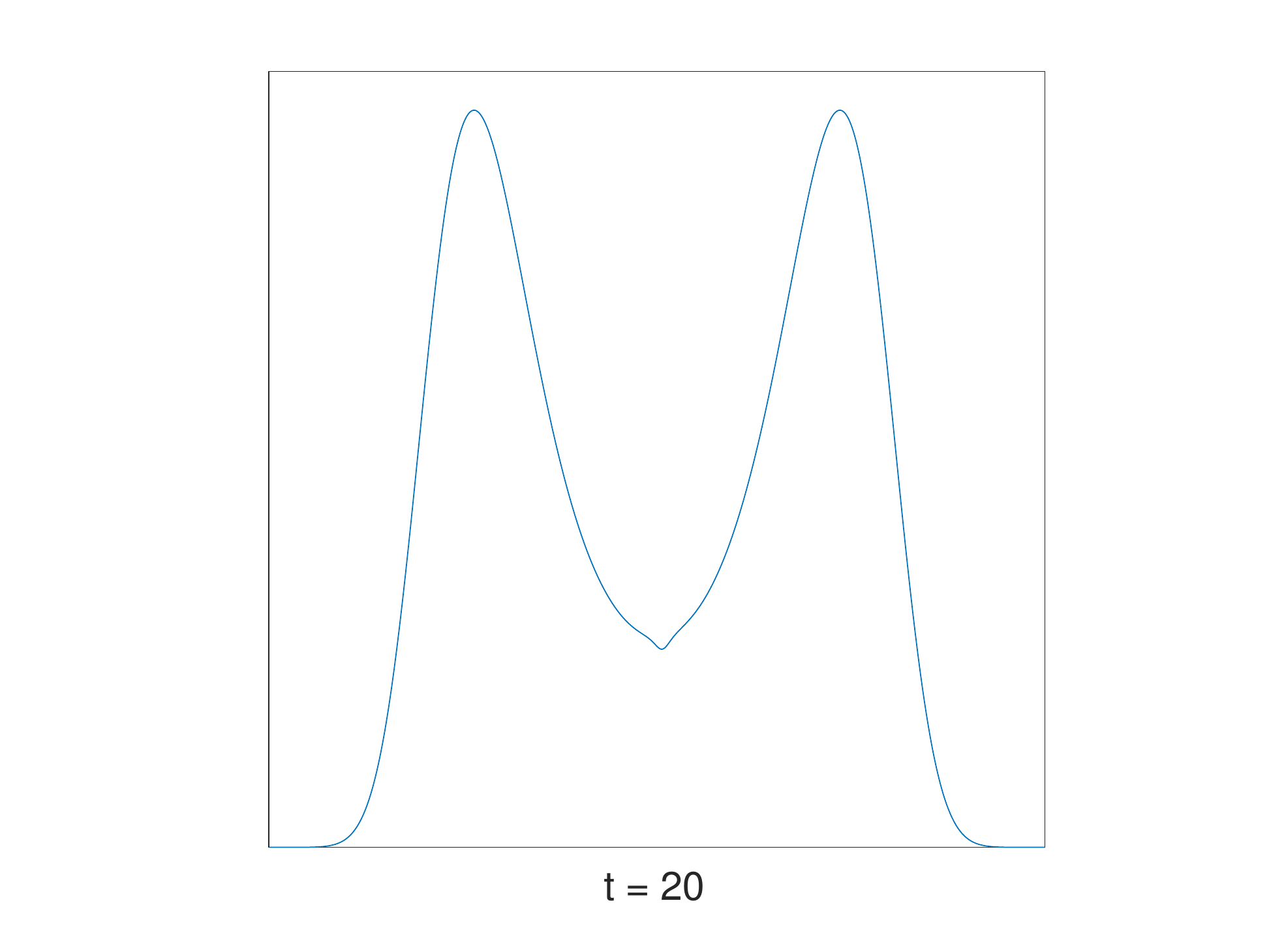}\includegraphics[bb=100bp 35bp 480bp 420bp,clip,scale=0.19]{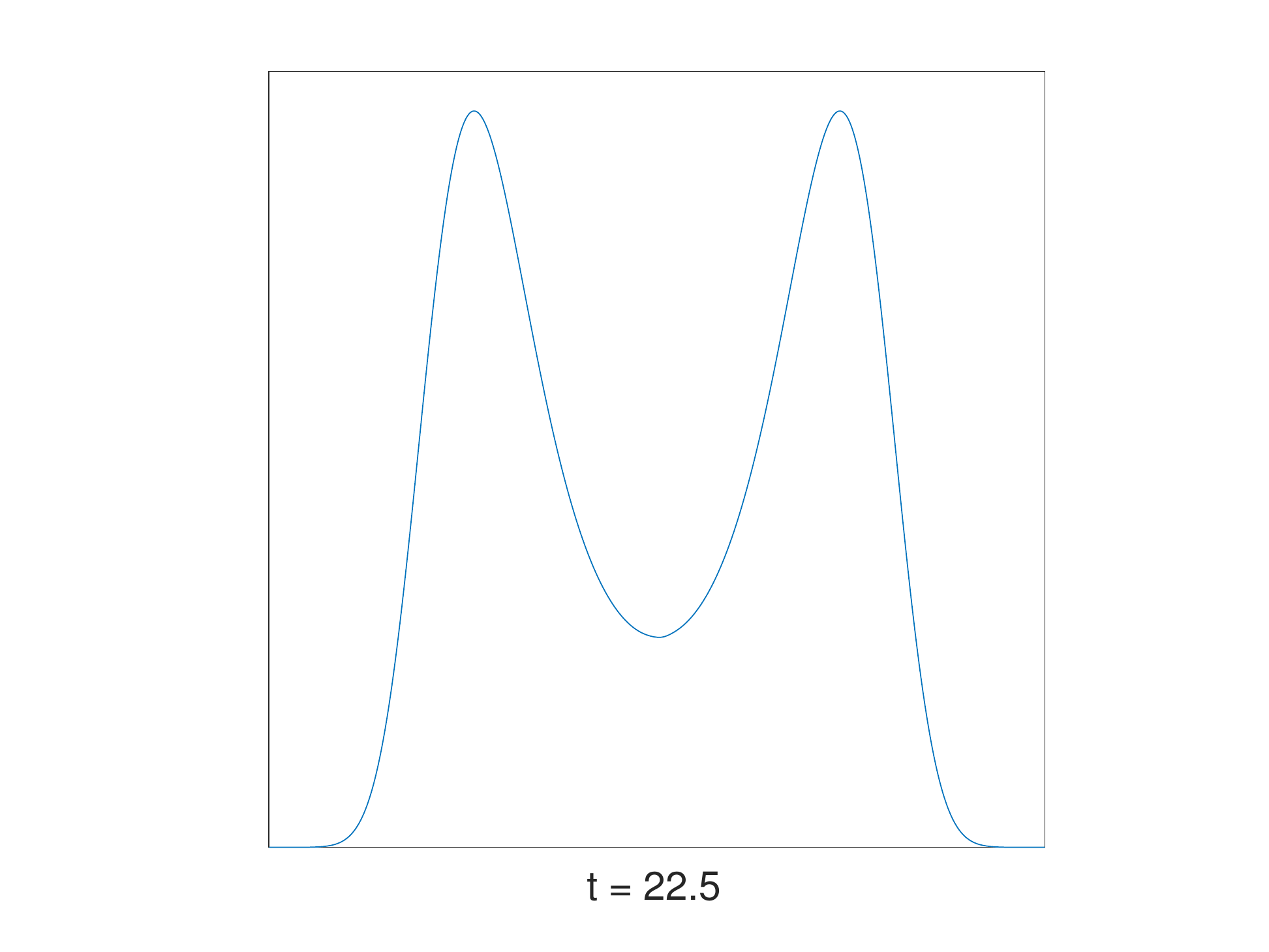}\caption{$\rho_{t}$ according to the nonlinear dynamics (\ref{eq:rhodot})
with $\beta=5$, $\sigma=0.0125$ at times $t=0,\,2.5,\,5,\,7.5,\,10,\,12.5,\,15,\,17.5,\,20\,,22.5$,
ordered left-to-right, then bottom-to-top. The profile at the last
frame ($t=22.5$) is visually indistinguishable from that of $\pi$. The interval of the horizontal axis is fixed as $[-1.5,1.5]$ in all figures, but the interval of the vertical axis varies to accommodate the changing vertical scale.}
\end{figure}

For the nonlinear dynamics when $\beta$ is large and $\sigma$ is
small, we observe transient behavior before the asymptotic convergence
regime. This corresponds to the regime in which the effective support
of $\rho$ expands to match that of $\pi$, at which point rapid convergence
ensues. This interpretation is visualized in Figure \ref{fig:movie}.

\begin{figure}
\centering{}\label{fig:metropolis}\includegraphics[bb=60bp 0bp 500bp 420bp,clip,scale=0.5]{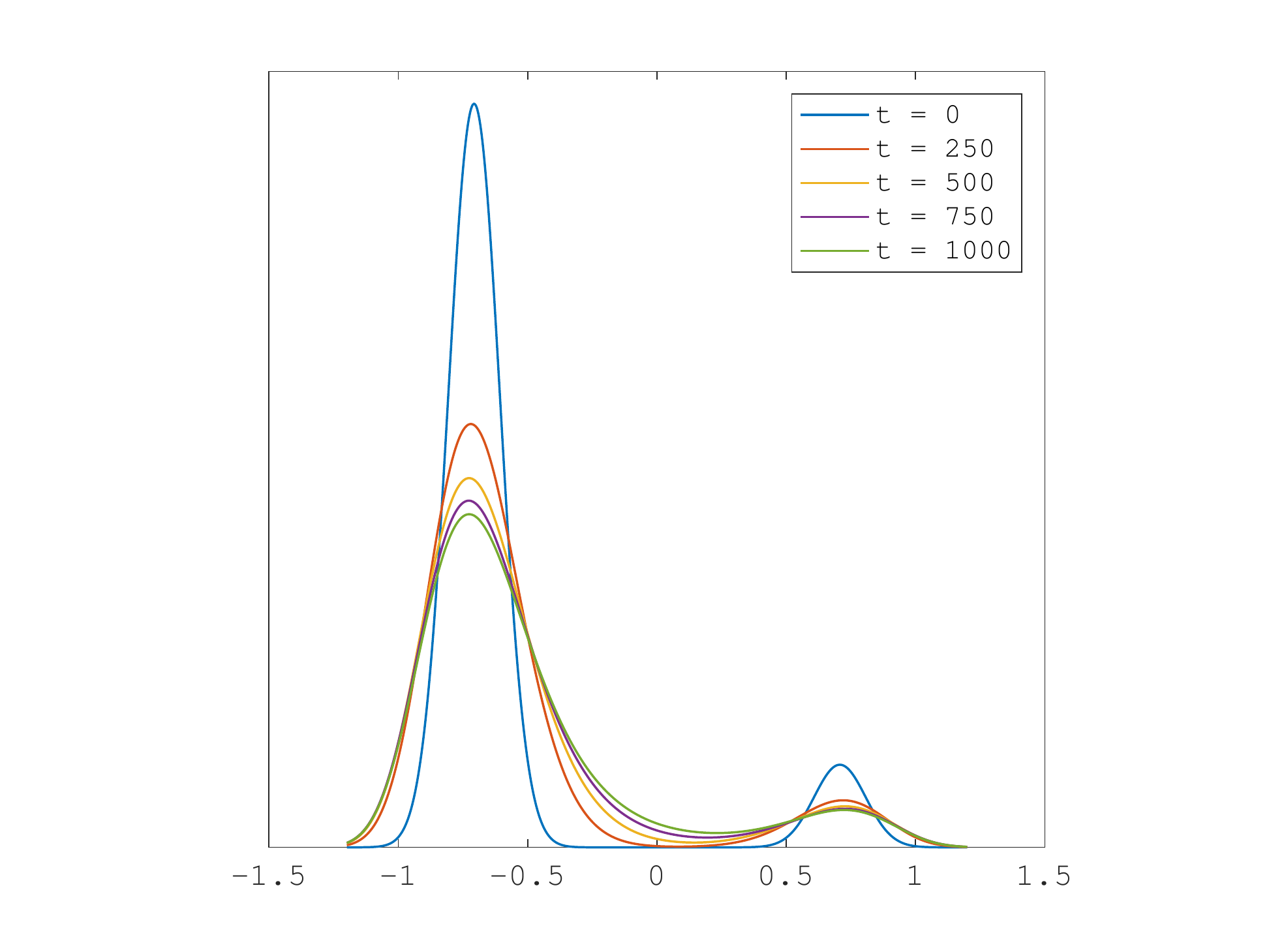}\caption{$\rho_{t}$ according to the continuum Metropolis dynamics (\ref{eq:contMetro})
with $\beta=5$, $\sigma=0.0125$ at several different times. Note
that even by time $t=1000$, the dynamics are far from convergence,
and the height of the second mode has actually decreased relative
to the initial condition.}
\end{figure}

Observe that even in the pre-asymptotic regime, the dynamics are able to achieve approximate balance between the probabilities of the two modes.
This behavior (which may be viewed as arising from the nonlocal walker
moves in the underlying ensemble scheme) contrasts sharply
with that of the continuum Metropolis dynamics (\ref{eq:contMetro}) for the same
problem, visualized in Figure \ref{fig:metropolis}. Those dynamics can be viewed as locally `bulldozing' probability from left to right, and in fact the
height of the second mode initially decreases.

\subsection{Gaussian progress regression with Bayesian hyperparameters}\label{sec:gpr}

In this section we consider the application of our method to Bayesian
inference of hyperparameters in Gaussian process regression.
For consistency with the application, the variable names in this section
are not consistent with the choices made for the general setting considered
above. The example problems are adapted from one considered in \cite{stacking},
which is also concerned with sampling for multimodal distributions.

In our experiments, we assess the efficiency of our methods in terms of integrated autocorrelation times (IAT) \cite{Sokal1997}. We are especially interested in the dependence of the efficiency on the number $N$ of walkers, with the case $N=1$ corresponding to an ordinary chain.

Specifically, we compute the average of one of the hyperparameters over the ensemble of walkers at each time to produce a time series. We define one step to be a move of a single walker. For an ensemble of $N$ walkers, we multiply the IAT of the aforementioned time series (estimated via the {\tt emcee} software package \cite{Foreman2013}) by a factor of $1/N$. This allows for a fair comparison between different ensemble sizes. To see this, consider an ensemble scheme with $N$ walkers which do not interact. The dynamics should be identical to $N$ independent chains, each with a single walker. Since one step is defined by a move of one walker, we will need $N$ steps to move each independent chain once. Thus, dividing the IAT by $N$ makes the result consistent with that of a single chain. Note, moreover, that in an efficient implementation, the computational cost of our method (as measured by the number of calls to the likelihood function) with $N$ interacting walkers is equivalent to the cost of running $N$ non-interacting chains.  For more on measuring convergence of ensemble schemes see \cite{GoodmanWeare:2010:EnsembleMCMC}.

\subsubsection{Univariate case}

First we consider a univariate mean-zero Gaussian process $\mathcal{GP}\left(0,\Sigma\right)$;
see Appendix \ref{app:gp} for relevant background. We take the covariance
to be 
\[
\Sigma(x_{1},x_{2})=\alpha^{2}\exp\left(-\frac{(x_{1}-x_{2})^{2}}{\rho^{2}}\right),
\]
 where $\alpha$ and $\rho$ are parameters (that we want to infer).
These parameters, if known, specify our prior distribution $\mathcal{GP}\left(0,\Sigma\right)$
for an unknown function $f$.

Let us also assume that we are given several $x_{i}$, $i=1,\ldots,m$
and that we have observed the function values at these points, corrupted
by some Gaussian noise, i.e., we have observed the data 
\[
y_{i}=f(x_{i})+\epsilon_{i},
\]
 where $\epsilon_{i}\sim\mathcal{N}(0,\sigma^{2})$. Here $\sigma$
is another model parameter that we wish to infer.

Let us collect our parameters as $\theta=(\alpha,\rho,\sigma)$ and set  $f_\mathbf{x} = (f(x_1),\dots,f(x_m))$. Fix
$K_{\theta}:=K(\mathbf{x},\mathbf{x})$, defined as in Appendix \ref{app:gp},
where here the subscript indicates the dependence of $K$ on $\theta$.
Then note that 
\[
y=f_{\mathbf{x}}+\epsilon
\]
 is a sum of independent Gaussians with distributions $\mathcal{N}(0,K)$
and $\mathcal{N}(0,\sigma^{2}I)$. Hence $y$ is distributed
as $\mathcal{N}(0,K+\sigma^{2}I)$.

Let $p(\theta)$ denote our prior for $\theta$. We seek to sample
$\theta$ according to 
\[
p(\theta\,\vert\,y)\propto p(y\,\vert\,\theta)p(\theta)\propto\vert K_{\theta}+\sigma^{2}I\vert^{-1/2}\,e^{-\frac{1}{2}y^{\top}\left(K_{\theta}+\sigma^{2}I\right)^{-1}y}\,p(\theta),
\]
 where $y$ is fixed throughout.

For our experiments, we choose independent $\mathrm{Cauchy}^{+}(0,3)$
priors for $\theta=(\alpha,\rho,\sigma)$. Moreover we generate data
$\mathbf{x}$ according to $x_i\sim\mathcal{N}(0,1)$ and $y$ according to $y_{i}=f_{\mathrm{true}}(x_{i})+\delta_{i}$,
where 
\begin{equation}
    f_{\mathrm{true}}(x_i) = 0.3 + 0.4x_i + 0.5\sin(2.7x_i) + 1.1 / (1+x_i^2)
    \label{eq:ftrue}
\end{equation}
and 
\begin{equation}
    \delta_i \sim
    \begin{cases}
    	\mathcal{N}(0, 0.125^2) & |x_i|< 1.5\\
        \mathcal{N}(0, 1.25^2) & \text{otherwise}.
    \end{cases}
    \label{eq:delta}
\end{equation}

\begin{figure}
\label{fig:post1}
\centering\includegraphics[scale=0.32]{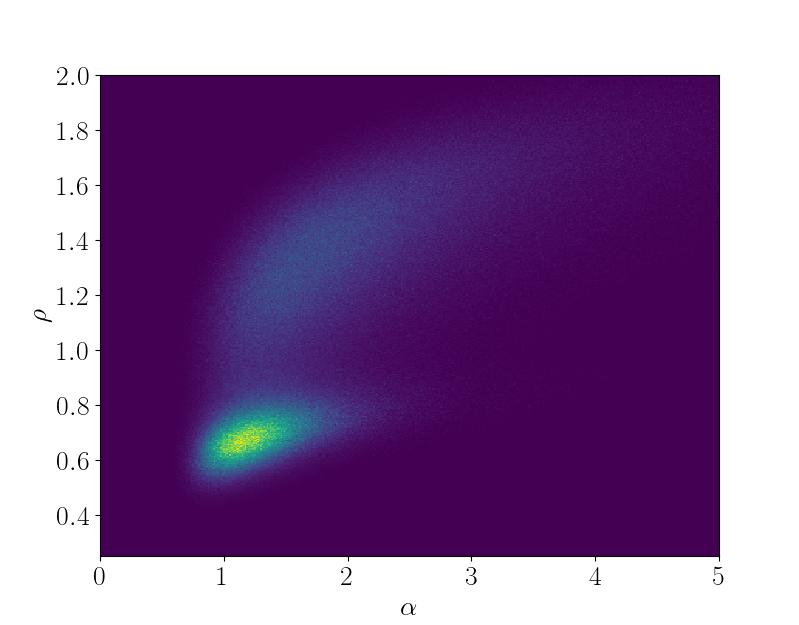}\includegraphics[scale=0.32]{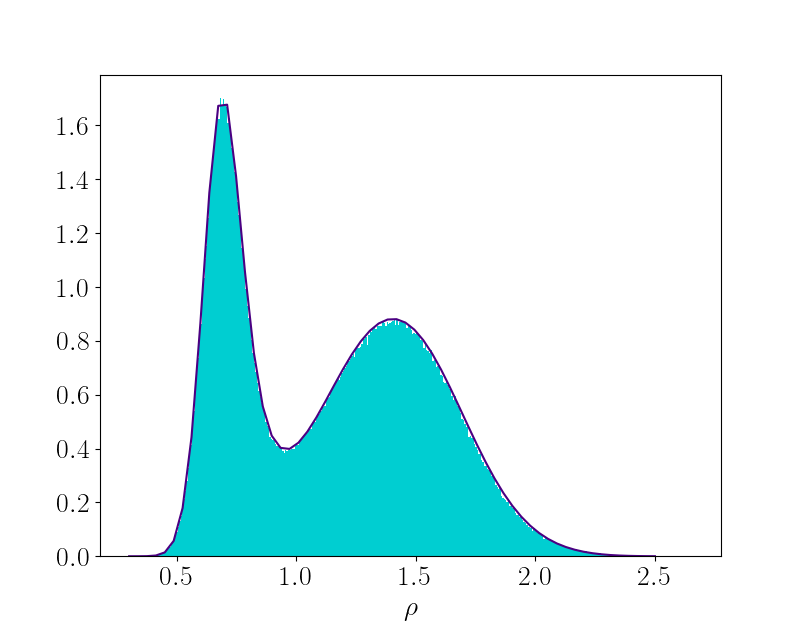}\caption{Univariate case. Posterior marginal distribution of $\alpha$ and $\rho$ (left) and numerically integrated density compared with sampled posterior of $\rho$ (right), obtained with ensemble size $N=50$.}
\end{figure}

\begin{figure}
\label{fig:AT1}
\centering\includegraphics[scale=0.32]{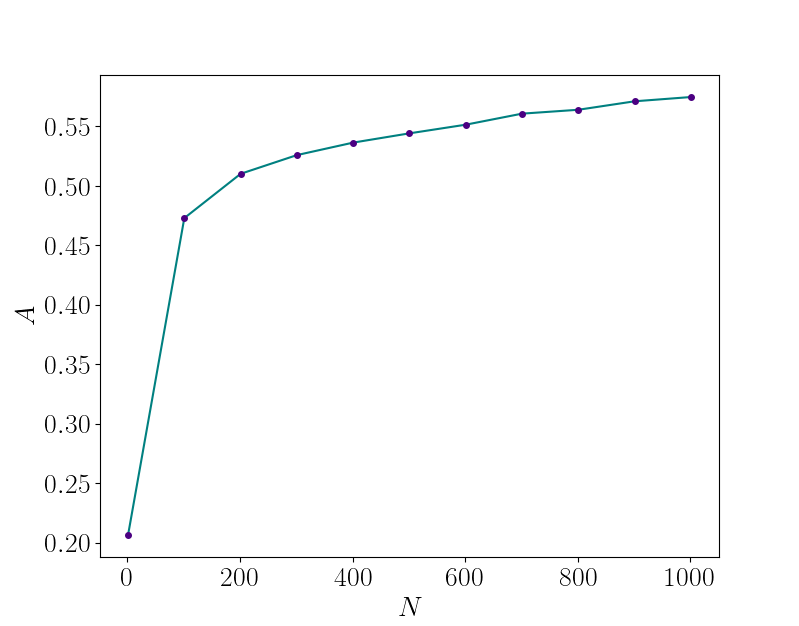}\includegraphics[scale=0.32]{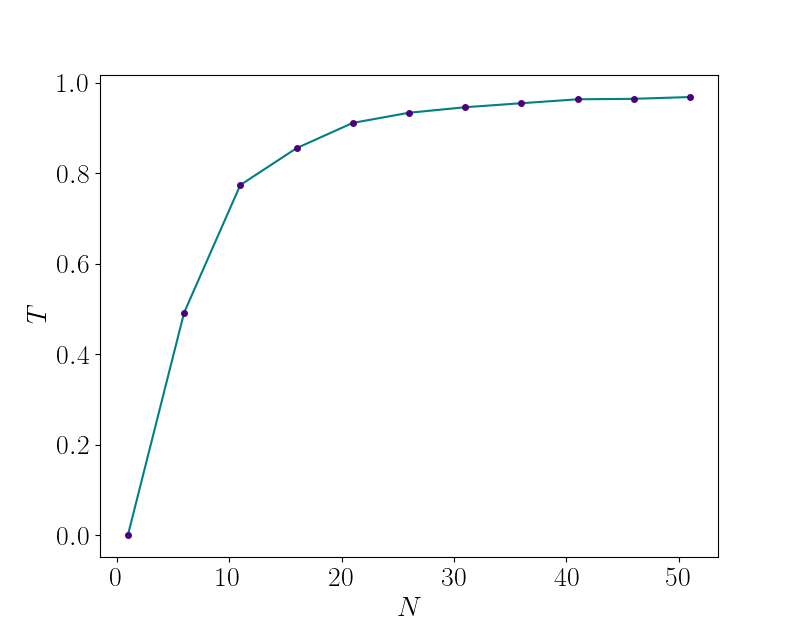}\caption{Univariate case. The acceptance probability $A$ versus $N$ (left) and the teleport probability $T$ versus $N$ (right), where $N$ is the number of walkers.}
\end{figure}

We sample from $p(\theta \, \vert \, y)$ using the ensemble method of Section \ref{sec:proposal}, where the proposal $q(\,\cdot\, \vert \theta)$ is $\mathcal{N}(\theta,\beta^2 I)$, $\beta^2 = 0.01$. In Figure \ref{fig:post1}, we plot posterior marginal distributions estimated from samples and compare against a ground truth obtained via numerical quadrature, which is feasible since $\theta$ is only 3-dimensional. Notice the multimodality of these marginals, suggesting the possibility of an advantage for the interacting walker scheme.
In Table \ref{table:1}, we record estimated IATs for different ensemble sizes $N$, confirming the advantage of taking $N \gg 1$. In Figure \ref{fig:AT1}, we plot the empirical acceptance probability $A$ and empirical teleport probability $T$ as functions of $N$. (The teleport probability is the probability that the indices of the cloned and removed walkers are different.)

\begin{table}
    \centering\begin{tabular}{| c || c | c | c|}
        \hline
        $N$ & 1 & 10 & 50 \\
        \hline
        IAT & 2111 & 857 & 97 \\ 
        \hline
    \end{tabular}
    \caption{Univariate case. Integrated autocorrelation times of the average of $\rho$ over all walkers.}
    \label{table:1}
\end{table}

\subsubsection{Multivariate case}

Next we consider the case of a multivariate Gaussian
process prior $\mathcal{GP}\left(0,\Sigma\right)$, where we take
\[
\Sigma(x_{1},x_{2})=\alpha^{2}\exp\left(-(x_{1}-x_{2})^{\top}ZZ^{\top}(x_{1}-x_{2})\right).
\]
 Here $\alpha\in\R$ and $Z\in\R^{n\times n}$ (upper triangular)
are parameters that we want to infer. Accordingly we collect our hyperparameters as $\theta=(\alpha,Z,\sigma)$.
We maintain the same priors on $\alpha$ and $\sigma$, but we must
specify a special prior for the upper triangular hyperparameter $Z$.

We want to choose a prior for $Z$ such that $ZZ^{\top}$ is distributed
according to $W_{n}(I_{n},n)$, which is the Wishart distribution
\cite{introMultivariate} with $n$ degrees of freedom and scale matrix $I_{n}$.
Following the Bartlett decomposition \cite{introMultivariate}, $Z$ is sampled as
\[
Z=\left(\begin{array}{cccc}
c_{1} & 0 & \cdots & 0\\
z_{21} & c_{2} & \cdots & 0\\
\vdots & \vdots & \ddots & \vdots\\
z_{n1} & z_{n2} & \cdots & c_{n}
\end{array}\right),
\]
 where the entries are all independently distributed, $z_{ij}\sim\mathcal{N}(0,1)$
for all $i>j$, and $c_{i}$ is distributed according to the chi distribution\emph{
}with $n-i+1$ degrees of freedom.

We generate data $\mathbf{x}$ according to $x_{i}\sim\mathcal{N}(0,I_n)$ and $y$
according to $y_{i}=f_{\mathrm{true}}(x_{i})+\delta_{i}$, where $$f_{\mathrm{true}}(x_i) = \prod_{j=1}^n \left(0.3 + 0.4x_{ij} + 0.5\sin(2.7x_{ij}) + 1.1 / (1+x_{ij}^2)\right)$$ and $\delta_i = \sum_{j=1}^n \delta_{ij}$, where 
$$\delta_{ij} \sim
\begin{cases}
	\mathcal{N}(0, 0.125^2) & |x_{ij}|< 1.5 \\
    \mathcal{N}(0, 1.25^2) & \text{otherwise}.
\end{cases}$$
For our experiment we fix $n=3$.

Again we sample from $p(\theta \, \vert \, y)$ using the ensemble method of Section \ref{sec:proposal}, where the proposal $q(\,\cdot\, \vert \theta)$ is $\mathcal{N}(\theta,D)$, 
\[
D = \left(\begin{array}{ccc}
    0.1 & 0 & 0 \\
    0 & 0.01 & 0 \\
    0 & 0 & 0.01
\end{array}\right).
\]

\begin{figure}
\label{fig:post2}
\centering\includegraphics[scale=0.32]{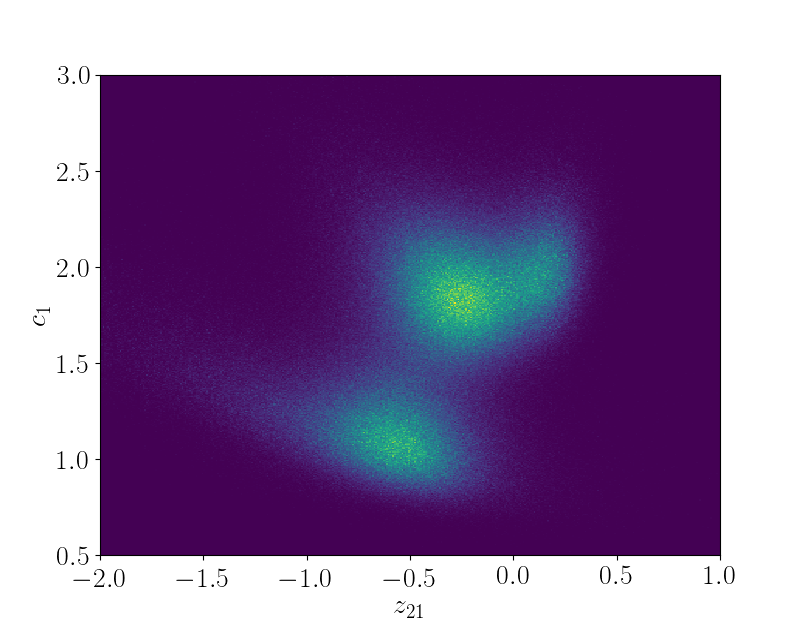}\includegraphics[scale=0.32]{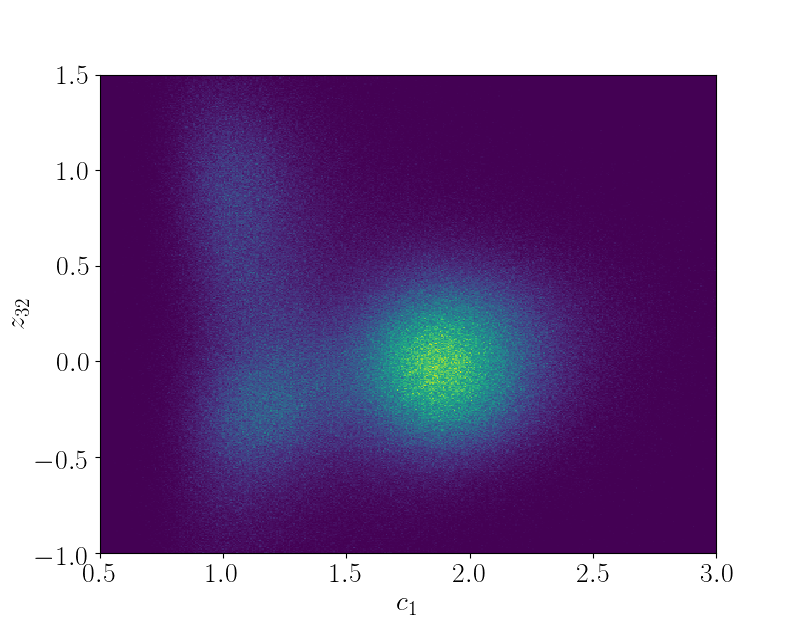}\caption{Multivariate case, $n=3$. Posterior marginal distribution of $z_{21}$ and $c_1$ (left) and of $c_1$ and $z_{32}$ (right), obtained with ensemble size $N=100$.}
\end{figure}

\begin{figure}
\label{fig:post3}
\centering
\includegraphics[scale=0.5]{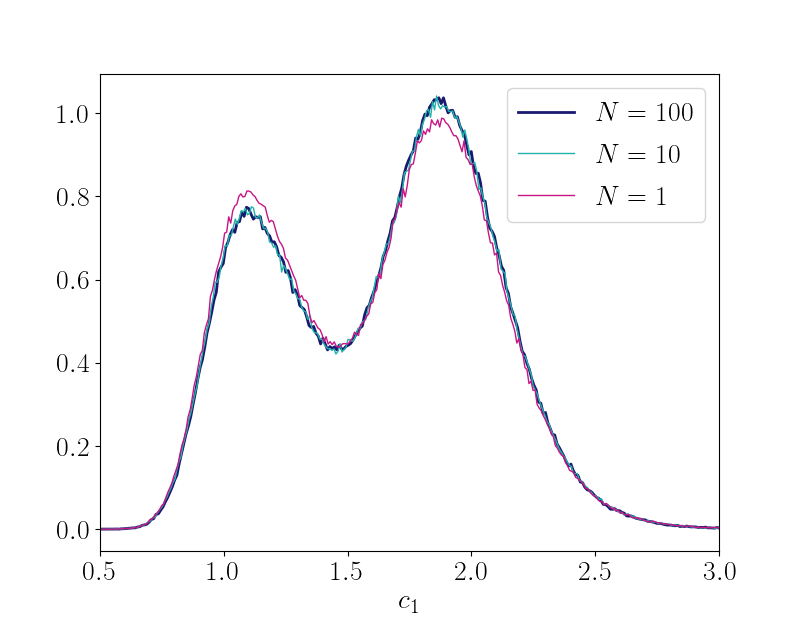}\caption{Multivariate case, $n=3$. Sampled posterior marginal distribution of $c_1$, obtained with ensemble sizes $N=1$, $10$, and $100$ and $10^7$ time steps. For $N>1$, only one walker (specifically, the one that was cloned) was entered into the histogram per step, so using the same number of time steps for each ensemble size is a fair comparison. Note the visible discrepancy for $N=1$ due to a long autocorrelation time.}
\end{figure}

In Figures \ref{fig:post2} and \ref{fig:post3}, we plot posterior marginal distributions estimated from samples, though now validation via numerical quadrature is not feasible due to the increased dimension of $\theta$. Again we observe multimodality, and Table \ref{table:2} demonstrates improved efficiency for large ensemble sizes $N$.

\begin{table}
    \centering\begin{tabular}{| c || c | c | c | c | c |}
        \hline
        $N$ & 1 & 10 & 20 & 50 & 100 \\
        \hline
        IAT & 1309 & 461 & 292 & 145 & 81 \\ 
        \hline
    \end{tabular}
    \caption{Multivariate case, $n=3$. Integrated autocorrelation times of the average of $c_1$ over all walkers.}
    \label{table:2}
\end{table}

\subsubsection{Non-Gaussian noise model}

Finally we return to the univariate case but consider a non-Gaussian
noise model for the $\epsilon_{i}$. Note that in this general case,
we cannot explicitly `integrate out' the $\epsilon_{i}$ as above,
and we are forced to think of them as additional Bayesian parameters
to be sampled. Then we must consider an expanded prior $p(\theta,\epsilon)=p(\theta)g_{\theta}(\epsilon)$,
where $g_{\theta}$ denotes our non-Gaussian noise model, which may
itself depend on the hyperparameters $\theta$. Then we want to sample
$\theta,\epsilon$ according to 
\[
p(\theta,\epsilon\,\vert\,y)\propto p(y\,\vert\,\theta,\epsilon)p(\theta)g_{\theta}(\epsilon)\propto\vert K_{\theta}\vert^{-1/2}\,e^{-\frac{1}{2}(y-\epsilon)^{\top}K_{\theta}^{-1}(y-\epsilon)}\,p(\theta)\,g_{\theta}(\epsilon),
\]
 where $y$ is fixed throughout. Since $K_{\theta}$ is usually numerically
low-rank, this expression is not suitable for sampling. We consider
the change of variable $(\theta,\epsilon)\ra(\theta,w)$ defined by
$\epsilon=y+K_{\theta}^{1/2}w$, motivating us to sample $\theta,w$
according to 
\[
p(\theta,w\,\vert\,y)\propto e^{-\frac{1}{2}\Vert w\Vert^{2}}\,p(\theta)\,g_{\theta}(\epsilon).
\]
 We take the same prior $p(\theta)$ for $\theta=(\alpha,\rho,\sigma)$
as above, and for our noise prior we consider independent Student-$t$
distributions for each $\epsilon_{i}$, each with mean $0$, scale
$\sigma$ (a hyperparameter), and $\nu=2$ degrees of freedom.

We generate data $\mathbf{x}$ according to $x_i\sim\mathcal{N}(0,1)$ and $y$
according to $y_{i}=f_{\mathrm{true}}(x_{i})+\delta_{i}$, where $f_{\mathrm{true}}$ and $\delta_i$ are the same as in \eqref{eq:ftrue} and \eqref{eq:delta}.

We sample from $p(\theta, w \, \vert \, y)$ using the method of Section \ref{sec:subset}, employing walker interaction only for the $\theta$ variables. The proposals $q(\,\cdot\, \vert \theta)$ and $r(\,\cdot\, \vert w)$ are distributed according to $\mathcal{N}(\theta,D)$ and $\mathcal{N}(w,\beta^2 I)$, respectively, where 
\[
D = \left(
\begin{array}{ccc}
    0.001 & 0 & 0 \\
    0 & 0.001 & 0 \\
    0 & 0 & 0.0001
\end{array}
\right)
\]
and $\beta^2 = 0.001$. We run the parallel chains for the $w$ variables for $30$ steps between each update step for the interacting $\theta$ variables. 
In Figure \ref{fig:post3}, we plot a posterior marginal distribution estimated from samples. Notice that, relative to Figure \ref{fig:post1}, the previously observed multimodality vanishes for this noise model. Nonetheless, we still see an advantage for large ensembles in Table \ref{table:3}.

\begin{figure}
\label{fig:post4}
\centering\includegraphics[scale=0.5]{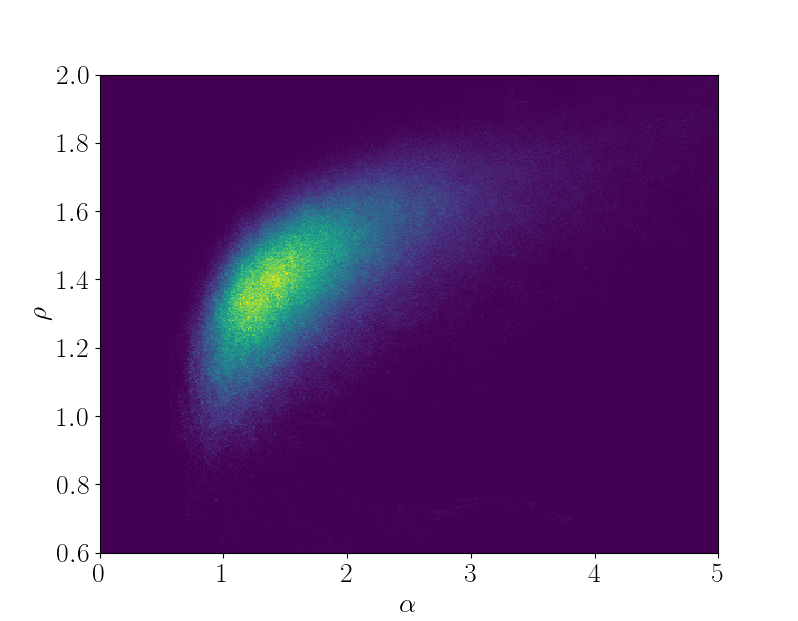}\caption{Non-Gaussian noise model. Posterior distribution of $\alpha$ and $\rho$, obtained with ensemble size $N=60$. Note that with the Student-$t$ noise model, we lose the multimodality in $\rho$.}
\end{figure}

\begin{table}
    \centering\begin{tabular}{| c || c | c | c | c |}
        \hline
        $N$ & 1 & 20 & 40 & 60 \\
        \hline
        IAT & 26016 & 20453 & 12428 & 6090 \\ 
        \hline
    \end{tabular}
    \caption{Non-Gaussian noise model. Integrated autocorrelation times of the average of $\rho$ over all walkers.}
    \label{table:3}
\end{table}

\section*{Acknowledgments}
We thank Omiros Papaspiliopoulos and Timoth\'ee Stumpf-F\'etizon for their help specifying the Gaussian process regression test problems in this paper.

\appendix

\section{Acceptance probability computations\label{app:acceptance}}

Observe that the likelihood $Q(\mathbf{x}'\,\vert\,\mathbf{x})$ of
the proposal of Section \ref{sec:proposal} is given by 
\[
Q(\mathbf{x}'\,\vert\,\mathbf{x})=\begin{cases}
w_{i}(\mathbf{x},x_{i}')\frac{1}{N}\sum_{k=1}^{N}q(x_{i}'\vert x_{k}), & \mbox{if \ensuremath{\mathbf{x}'} and \ensuremath{\mathbf{x}} differ on a unique index \ensuremath{i},}\\
0, & \mbox{otherwise}.
\end{cases}
\]
Supposing that we have generated $\mathbf{x'}$ via the procedure
described in Section \ref{sec:proposal} (i.e., so that $i$, $j$,
and $z$ are defined as above), the Metropolis-Hastings acceptance
probability is given by 
\begin{eqnarray*}
A & = & \min\left(1,\frac{\Pi(\mathbf{x}')}{\Pi(\mathbf{x})}\frac{Q(\mathbf{x}\,\vert\,\mathbf{x}')}{Q(\mathbf{x}'\,\vert\,\mathbf{x})}\right) \\
 & = & 
 \min\left(1,\frac{\pi(z)}{\pi(x_{i})}\frac{w_{i}(\mathbf{x}',x_{i})}{w_{i}(\mathbf{x},z)}\frac{\sum_{k=1}^{N}q(x_{i}\vert x_{k}')}{\sum_{k=1}^{N}q(z\vert x_{k})}\right)\\
 & = & \min\left(1,\frac{\pi(z)}{\pi(x_{i})}\frac{Z(\mathbf{x},z)}{Z(\mathbf{x}',x_{i})}\frac{\left(\frac{q(x_{i}'\,\vert\,x_{i})+\sum_{k\neq i}^{N}q(x_{i}'\,\vert\,x_{k}')}{\pi(x_{i}')}\right)}{\left(\frac{q(x_{i}\,\vert\,z)+\sum_{k\neq i}^{N}q(x_{i}\,\vert\,x_{k})}{\pi(x_{i})}\right)}\frac{\sum_{k=1}^{N}q(x_{i}\,\vert\,x_{k}')}{\sum_{k=1}^{N}q(z\,\vert\,x_{k})}\right).
\end{eqnarray*}
 But recall $x_{i}'=z$, and $x_{k}'=x_{k}$ for $k\neq i$, so 
\begin{eqnarray*}
A & = & \min\left(1,\frac{Z(\mathbf{x},z)}{Z(\mathbf{x}',x_{i})}\frac{q(z\,\vert\,x_{i})+\sum_{k\neq i}^{N}q(z\,\vert\,x_{k})}{q(x_{i}\,\vert\,z)+\sum_{k\neq i}^{N}q(x_{i}\,\vert\,x_{k})}\frac{\sum_{k=1}^{N}q(x_{i}\vert x_{k}')}{\sum_{k=1}^{N}q(z\vert x_{k})}\right)\\
 & = & \min\left(1,\frac{Z(\mathbf{x},z)}{Z(\mathbf{x}',x_{i})}\frac{\sum_{k=1}^{N}q(z\,\vert\,x_{k})}{\sum_{k=1}^{N}q(x_{i}\,\vert\,x_{k}')}\frac{\sum_{k=1}^{N}q(x_{i}\,\vert\,x_{k}')}{\sum_{k=1}^{N}q(z\,\vert\,x_{k})}\right)\\
 & = & \min\left(1,\frac{Z(\mathbf{x},z)}{Z(\mathbf{x}',x_{i})}\right),
\end{eqnarray*}
 as desired.

\section{Global convergence proof\label{app:global}}

\begin{proof}
For consistency of presentation, we will maintain the continuous notation,
i.e., writing integrals over $X$ instead of sums.

From the dynamics \eqref{eq:rhodot} we have 
\[
\dot{\rho}=\mathcal{Q}\rho-Z_{\rho}^{-1}\frac{\mathcal{Q}\rho}{\pi}\rho =: G[\rho].
\]
Note that $G[\rho](x) > 0$ if $\rho(x) = 0$ because $\mathcal{Q} \rho$ has full support. By the  continuity of $G$ and the compactness of the space of probability measures, for any $x$, we have $\dot{\rho}(x) = G[\rho](x) > 0$ if $\rho (x) < \delta$ for some $\delta > 0$ sufficiently small. Consequently $\mathrm{supp}(\rho_{t})=X$ for all $t>0$ at which
$\rho_{t}$ is defined (even if $\mathrm{supp}(\rho_{0})\neq X$).
Moreover, as the constraint that $\int\rho\,dx=1$ is conserved by
the dynamics \eqref{eq:rhodot}, we also have that $\rho_{t}$ lies within
the probability simplex for all times $t$ at which it is defined.
This \emph{a priori }bound within a compact region, together with
a Lipschitz condition on the dynamics within this domain, guarantees
global-in-time existence of $\rho_{t}$ by standard theory (cf., \cite{TeschlODE}).

Recall \eqref{eq:monotone}, i.e., that
\[
\frac{d}{dt}\chi^{2}(\pi\, \Vert\, \rho_{t}) \leq - \mathrm{Var}_{\mathcal{Q}\rho}(\pi/\rho_t),
\]
Define the sublevel set 
 \[
 S_{b} := \left\{ \rho\ \mathrm{prob.\,dens.} \,:\,\chi^{2}(\pi\,\Vert\,\rho)\leq b \right\}, 
 \]
and note by monotonicity that setting $b= \chi^2 (\pi\,\Vert\,\rho_{0})$, we have $\rho_t \in S_{b}$ for all $t$. Then evidently 
\[
\frac{d}{dt}\chi^{2}(\pi\, \Vert\, \rho_{t}) \leq - \mathrm{Var}_{\rho}(\pi/\rho_t) \inf_{\rho \in S_b} \left\{ \frac{\mathrm{Var}_{\mathcal{Q}\rho}(\pi/\rho)}{\mathrm{Var}_{\rho}(\pi/\rho)} \right\} = - \gamma^{-1}\, \chi^2( \pi \,\Vert \, \rho_t ),
\]
where $\alpha$ is defined as in the statement of the theorem. Then \eqref{eq:convbound} follows from Grönwall's inequality, provided we can show that $\gamma < +\infty$. Note that $\gamma < +\infty$ holds if we can show \eqref{eq:infbound}, so remains only to show \eqref{eq:infbound}.

Now 
\begin{eqnarray*}
\mathrm{Var}_{\mathcal{Q}\rho}\left[\pi/\rho\right] & = & \int\left[\frac{\pi}{\rho}-\left(\int\frac{\pi}{\rho}\,\mathcal{Q}\rho\,dx\right)\right]^{2}\,\mathcal{Q}\rho\,dx\\
 & \geq & \Vert\rho/\mathcal{Q}\rho\Vert_{\infty}^{-1}\int\left[\frac{\pi}{\rho}-\left(\int\frac{\pi}{\rho}\,\mathcal{Q}\rho\,dx\right)\right]^{2}\,\rho\,dx.
\end{eqnarray*}
 But note that $\int\left(\frac{\pi}{\rho}-a\right)^{2}\rho\,dx$
is minimized over $a \in \mathbb{R}$ by taking $a=\int\frac{\pi}{\rho}\,\rho\,dx=1$, so 
\[
\int\left[\frac{\pi}{\rho}-\left(\int\frac{\pi}{\rho}\,\mathcal{Q}\rho\,dx\right)\right]^{2}\,\rho\,dx\ge\mathrm{Var}_{\rho}(\pi/\rho).
\]
 Hence $\mathrm{Var}_{\mathcal{Q}\rho}\left[\pi/\rho\right]\geq\Vert\rho/\mathcal{Q}\rho\Vert_{\infty}^{-1}\mathrm{Var}_{\rho}(\pi/\rho)$,
which implies \eqref{eq:infbound}.
\end{proof}

\section{Linearization computations and asymptotic convergence proof\label{app:linearization}}

Let 
\[
F(\eta)=\frac{1}{Z_{\pi+\eta}}\left[Z_{\pi+\eta}-\frac{\pi+\eta}{\pi}\right]\mathcal{Q}(\pi+\eta)
\]
 as Section \ref{sec:localconv}. Recall that $Z_{\rho}=\int\frac{\rho\mathcal{Q}\rho}{\pi}\,dx$.
In particular $Z_{\pi}=1$. We want to compute $DF(0)$. Now in our
expression for $F(\eta)$, the middle factor is zero when $\eta=0$,
hence in the product rule only one term contributes and we have 
\begin{eqnarray*}
\frac{\delta F(\eta)(x)}{\delta\eta(y)}\bigg\vert_{\eta=0} & = & \mathcal{Q}\pi(x)\frac{\delta}{\delta\eta(y)}\bigg\vert_{\eta=0}\left[Z_{\pi+\eta}-\frac{\pi(x)+\eta(x)}{\pi(x)}\right]\\
 & = & \mathcal{Q}\pi(x)\left[\frac{\delta}{\delta\eta(y)}\bigg\vert_{\eta=0}Z_{\pi+\eta}-\frac{\delta(x,y)}{\pi(x)}\right].
\end{eqnarray*}

To deal with the partition function, observe that 
\[
Z_{\rho}=\rho^{*}\left[\mathrm{diag}(\pi)^{-1}\mathcal{Q}\right]\rho=\frac{1}{2}\rho^{*}\left[\mathrm{diag}(\pi)^{-1}\mathcal{Q}+\mathcal{Q}^{*}\mathrm{diag}(\pi)^{-1}\right]\rho,
\]
 i.e., we may view $Z_{\rho}$ as a symmetric quadratic form in $\rho$.
Hence 
\[
\frac{\delta}{\delta\rho}Z_{\rho}=\left(\mathrm{diag}(\pi)^{-1}\mathcal{Q}+\mathcal{Q}^{*}\mathrm{diag}(\pi)^{-1}\right)\rho=\frac{\mathcal{Q}\rho}{\pi}+\mathcal{Q}^{*}\left(\frac{\rho}{\pi}\right).
\]
 But then 
\[
\frac{\delta}{\delta\eta(y)}\bigg\vert_{\eta=0}Z_{\pi+\eta}=\frac{\delta}{\delta\rho(y)}\bigg\vert_{\rho=\pi}Z_{\rho}=\frac{\mathcal{Q}\pi}{\pi}(y)+[\mathcal{Q}^{*}\mathbf{1}](y)=\frac{\mathcal{Q}\pi}{\pi}(y)+1,
\]
 where $\mathbf{1}$ is the constant function taking value $1$, and
we used that $\mathcal{Q}^{*}\mathbf{1}=1$ because $\mathcal{Q}$
is a Markov transition operator.

In summary we have established that
\[
\mathcal{J}(x,y):=\frac{\delta F(\eta)(x)}{\delta\eta(y)}\bigg\vert_{\eta=0}=\mathcal{Q}\pi(x)\left[\frac{\mathcal{Q}\pi}{\pi}(y)+1-\frac{\delta(x,y)}{\pi(x)}\right],
\]
 where $\mathcal{J}(x,y)$ denotes the kernel of the operator $DF(0)$.
Then 
\[
\mathcal{J}=\mathcal{Q}\pi\left(\frac{\mathcal{Q}\pi}{\pi}\right)^{*}-\mathrm{diag}\left(\frac{\mathcal{Q}\pi}{\pi}\right)+\left(\mathcal{Q}\pi\right)\mathbf{1}^{*}.
\]
 But since $\mathbf{1}^{*}\eta=0$ for all $\eta\in S$, we have that
\[
\mathcal{J}=\mathcal{Q}\pi\left(\frac{\mathcal{Q}\pi}{\pi}\right)^{*}-\mathrm{diag}\left(\frac{\mathcal{Q}\pi}{\pi}\right)
\]
 as an operator on $S$ (and indeed one verifies easily $S$ is invariant
under $\mathcal{J}$ so defined).

\begin{proof}[Proof of Theorem \ref{thm:continuum}]
For consistency of presentation, we maintain the continuous notation,
i.e., writing integrals over $X$ instead of sums. In the finite-dimensional
case, the computation of the Jacobian $DF(0)$ for the dynamics $\dot{\eta}=F(\eta)$
in Appendix \ref{app:linearization} is rigorous without further clarification.
Then standard stable manifold theory for ODEs (cf., Theorem 9.4 of
\cite{TeschlODE}) guarantees the result, provided we can show that
$\sigma(\mathcal{J})\subset\R$ with $\sup\sigma(\mathcal{J})<- \Vert \pi / \mathcal{Q} \pi \Vert_\infty^{-1}$.

First note that taking $\mathcal{D}:=\mathrm{diag}(\sqrt{\pi})$ we
have 
\[
\mathcal{M}:=\mathcal{D}^{-1}\mathcal{J}\mathcal{D}=\left(\frac{\mathcal{Q}\pi}{\sqrt{\pi}}\right)\left(\frac{\mathcal{Q}\pi}{\sqrt{\pi}}\right)^{*}-\mathrm{diag}\left(\frac{\mathcal{Q}\pi}{\pi}\right).
\]
 Then $\mathcal{M}$ is self-adjoint, hence diagonalizable with real
eigenvalues. Since $\mathcal{M}$ and $\mathcal{J}$ are similar,
$\mathcal{J}$ is also diagonalizable with the same eigenvalues. Note
that $\mathcal{M}$ is on operator on $\mathcal{D}^{-1}S=\{f\,:\,\int f\sqrt{\pi}\,dx=0\}$,
not on $S$.

To complete the proof it then suffices to show that $f^{*}\mathcal{M}f<-\Vert \pi / \mathcal{Q} \pi \Vert_\infty^{-1}\, f^{*}f$
for any $f$ with $\int f\sqrt{\pi}\,dx=0$. Observe that 
\begin{equation}
f^{*}\mathcal{M}f=\left(\int\frac{\mathcal{Q}\pi}{\sqrt{\pi}}f\,dx\right)^{2}-\int\frac{\mathcal{Q}\pi}{\pi}f^{2}\,dx.\label{eq:quadform}
\end{equation}
 Since $\int f\sqrt{\pi}\,dx=0$, we may write, for an arbitrary constant
$c$ (to be optimized later): 
\begin{eqnarray*}
\left(\int\frac{\mathcal{Q}\pi}{\sqrt{\pi}}f\,dx\right)^{2} & = & \left(\int\frac{\mathcal{Q}\pi-c\pi}{\sqrt{\pi}}f\,dx\right)^{2}\\
 & = & \left(\int\sqrt{\mathcal{Q}\pi}\,\frac{\sqrt{\mathcal{Q}\pi}-c\frac{\pi}{\sqrt{\mathcal{Q}\pi}}}{\sqrt{\pi}}f\,dx\right)^{2}\\
 & \leq & \left[\int\mathcal{Q}\pi\,dx\right]\left[\int\left(\sqrt{\mathcal{Q}\pi}-c\frac{\pi}{\sqrt{\mathcal{Q}\pi}}\right)^{2}\frac{f^{2}}{\pi}\,dx\right],
\end{eqnarray*}
 where the inequality follows from the Cauchy-Schwarz inequality.
But $\int\mathcal{Q}\pi\,dx=1$, and expanding the square in the other
integrand yields 
\[
\left(\int\frac{\mathcal{Q}\pi}{\sqrt{\pi}}f\,dx\right)^{2}\leq\int\frac{\mathcal{Q}\pi}{\pi}f^{2}\,dx-2c\int f^{2}\,dx+c^{2}\int\frac{\pi}{\mathcal{Q}\pi}f^{2}\,dx.
\]
By plugging into (\ref{eq:quadform}) we see that 
\[
f^{*}\mathcal{M}f\leq-2c\int f^{2}\,dx+c^{2}\int\frac{\pi}{\mathcal{Q}\pi}f^{2}\,dx.
\]
 Then we want to optimize this bound over $c$. Evidently the optimal
$c$ is given by 
\[
c=\frac{\int f^{2}\,dx}{\int\frac{\pi}{\mathcal{Q}\pi}f^{2}\,dx},
\]
 which yields 
\[
f^{*}\mathcal{M}f\leq-\frac{(f^{*}f)^{2}}{\int\frac{\pi}{\mathcal{Q}\pi}f^{2}\,dx}.
\]
 But $\int\frac{\pi}{\mathcal{Q}\pi}f^{2}\,dx\leq \Vert \pi / \mathcal{Q} \pi \Vert_\infty\, f^{*}f$,
so $f^{*}\mathcal{M}f\leq- \Vert \pi / \mathcal{Q} \pi \Vert_\infty^{-1}\, f^{*}f$, as was to be shown.
\end{proof}

\section{Gradient flow computations\label{app:gradflow}}

Expanding the expression in (\ref{eq:argmin}) to lowest order we
obtain the asymptotically equivalent problem: 
\[
\rho_{\ve}=\underset{\tilde{\rho}\in\mathcal{P}(X)}{\mbox{argmin}}\left\{ \int\frac{\delta E(\rho)}{\delta\rho(x)}\,(\tilde{\rho}(x)-\rho(x))\ dx+\frac{1}{8\ve}\int\frac{(\tilde{\rho}(x)-\rho(x))^{2}}{\rho(x)}\ dx\right\} .
\]
 Now 
\[
\frac{\delta}{\delta\rho(x)}E(\rho)=\frac{1}{8}\frac{\delta}{\delta\rho(x)}\int\left(1-\frac{\rho}{\pi}\right)^{2}\pi\,dx=\frac{1}{4}\left(\frac{\rho(x)}{\pi(x)}-1\right),
\]
 so we must solve 
\[
\underset{\tilde{\rho}\in\mathcal{P}(X)}{\mbox{argmin}}\left\{ \frac{1}{4}\int\left(\frac{\rho}{\pi}-1\right)(\tilde{\rho}-\rho)\ dx+\frac{1}{8\ve}\int\frac{(\tilde{\rho}-\rho)^{2}}{\rho}\ dx\right\} .
\]
for which the optimality condition is 
\[
1 - \frac{\rho}{\pi}=\frac{1}{\ve}\frac{\tilde{\rho}-\rho}{\rho}-\lambda,
\]
 where $\lambda$ is a constant, namely the Lagrange multiplier for
the constraint $\int\rho\,dx=1$. Rearranging we obtain 
\[
\rho_{\ve}=\rho+\ve\left[\left(1-\frac{\rho}{\pi}\right)\rho+\lambda\rho\right],
\]
 where $\lambda$ is chosen so that $\int\rho_{\ve}=1$. Notice that
this means precisely that $\lambda=C_{p}$, hence we obtain 
\[
\partial_{\tau}\rho=\left(1- \frac{\rho}{\pi}\right)\rho+C_{\rho}\,\rho,
\]
 as desired.

\section{Gaussian processes\label{app:gp}}

A Gaussian process is a random function $f:\R^{n}\ra\R$
specified by a mean $\mu(x)$ and covariance $\Sigma(x_{1},x_{2})$
which satisfy 
\[
\E\left[f(x)\right]=\mu(x)
\]
 and 
\[
\E\left[\left(f(x_{1})-\mu(x_{2})\right)\left(f(x_{1})-\mu(x_{2})\right)\right]=\Sigma(x_{1},x_{2}),
\]
 together with the specification that for any\emph{ }choice of $\mathbf{x}=(x_{1},\ldots,x_{m})\in\R^{n\times m}$,
the random vector 
\[
f_{\mathbf{x}}:=(f(x_{1}),\ldots,f(x_{n}))
\]
 is Gaussian distributed. Hence note that in particular $f_{\mathbf{x}}$
has mean 
\[
\left(\mu(x_{1}),\ldots,\mu(x_{n})\right)
\]
 and covariance 
\[
K(\mathbf{x},\mathbf{x}):=\left(\begin{array}{ccc}
\Sigma(x_{1},x_{1}) & \cdots & \Sigma(x_{1},x_{n})\\
\vdots & \ddots & \vdots\\
\Sigma(x_{n},x_{1}) & \cdots & \Sigma(x_{n},x_{n})
\end{array}\right).
\]
 In this case we say that $f\sim\mathcal{GP}\left(\mu,\Sigma\right)$.

\bibliographystyle{siamplain}
\bibliography{refs}
\end{document}